\documentclass{article}
\usepackage{algorithm}
\usepackage{algpseudocode}
\usepackage[utf8]{inputenc}
\usepackage{xcolor}
\usepackage{multicol} 
\usepackage[english]{babel}
\usepackage{booktabs}
\usepackage[letterpaper,top=2cm,bottom=2cm,left=3cm,right=3cm,marginparwidth=1.75cm]{geometry}
\usepackage{longtable}

\usepackage{amsmath}
\usepackage{graphicx}
\usepackage[colorlinks=true, allcolors=blue]{hyperref}

\title{A multiscale model of friction considering the influence of third-body wear particles}

\author{ P. Alavi$^*$, L. Rocchi$^+$, C. Leppin$^+$, J.F. Molinari$^*$, G. Anciaux$^*$ \\
$^{*}$
École Polytechnique Fédérale de Lausanne, EPFL, Switzerland \\
$^+$
Novelis R\&D Centre Sierre, Switzerland\\}

\date{}
\begin{document}

\newpage
\maketitle

\begin{abstract}
	Accurately predicting friction in sliding interfaces that contain third-body wear particles is critical for engineering applications such as sliding movement in pistons, bearings or metal forming. 
    We present a hierarchical multiscale framework that links particle-scale mechanics to macroscopic friction in a strip-draw friction test. At the macroscale, a one-dimensional finite-element model reproduces the global stress state of the strip-draw setup and updates the local wear-particle density via Archard’s law. The local friction force at each node is then computed from mesoscale simulation results. At the mesoscale, a coupled discrete-element boundary-element approach resolves load sharing between rough surfaces and rigid oblate-spheroidal wear particles. The mesoscale solution returns to the macroscale solver a friction coefficient that depends on normal pressure, sliding velocity, surface geometry, and particle density, thereby closing the loop between scales. The simulated friction coefficient matches strip-draw experiments, capturing both the observed decrease in friction with increasing normal pressure and the influence of tool-pad size.

\end{abstract}


\section{Introduction}
The coefficient of friction is used to quantify the resistance to relative sliding between contacting surfaces. It is defined as the ratio of the frictional force to the normal force \cite{Coulomb1785}. Despite its widespread use, this coefficient is inherently empirical and does not fully capture the complexities of frictional behavior \cite{Persson2000}.
Classical models, such as the Coulomb friction law, relate the friction force directly to the normal load, independent of contact area or sliding speed \cite{Coulomb1785}. Bowden and Tabor later showed that the friction coefficient is proportional to the real contact area that forms at the microscopic scale \cite{BowdenTabor1950}. However, numerous studies have demonstrated that the friction coefficient is not a fixed parameter and cannot be determined solely by surface roughness and applied load. Instead, it evolves with the sliding conditions, material interactions, and the contact history \cite{Rubino2017,Ronsin2000,Fortunato2015}.


One notable empirical framework that accounts for the evolution of contact conditions during sliding is the rate-and-state friction law, originally developed based on experimental observations by Dieterich and Ruina \cite{Dieterich1979,Ruina1983}. This model describes the interplay between the sliding rate and the evolving “state” of the contact interface, which encodes the contact history. Later, Marone \cite{Marone1998} synthesized nearly two decades of rock-friction experiments and systematized the rate-and-state friction parameters. Although originally developed to explain seismic fault behavior, the rate-and-state friction law has since been applied to other domains, such as tribology.
For instance, to model boundary-lubrication regime, Dupont and Dunlap \cite{Dupont1993} developed a phenomenological friction model tailored for precision drive systems operating under boundary lubrication. They introduced a single-state-variable framework in which the state variable represents the fraction of load supported by the adsorbed lubricant film. This model effectively captured the time-dependent frictional responses observed in lubricated steel contacts, at very low velocities.
Carlson and Batista \cite{Carlson1996} later proposed a constitutive relation to describe friction between surfaces separated by an atomically thin lubricant layer. Their model incorporated a state variable interpreted in terms of shear-induced structural transitions (shear melting) within the lubricant film.
%

In practice, other than the fluid lubricant, debris, dust, or solid-lubricant particles (so-called third bodies) are also entrained into the interface. These particles can carry load and alter the interface state during sliding, thereby influencing the friction coefficient \cite{colas2013,Mischler2013,Barlemont2023}. 
This indicates that the state variable could be redefined to incorporate the effects of third-body particles, an aspect that, to the best of our knowledge, appears to be underexplored in current studies.
The third body may differ chemically from the original contacting materials; however, even without chemical transformation, wear debris can play a fundamental mechanical role in tribological systems. Acting as an active medium, the third body continuously redistributes material and load at the contact interface, with debris evolving into a dynamic interfacial layer that governs friction and wear behavior \cite{Godet1984}. Denape \cite{Denape2015} illustrates, through various experiments and case studies, how the life cycle of wear debris, including its generation, movement, transformation, and eventual ejection, significantly influences friction and wear.
Colas et al. \cite{colas2013} experimentally tracked MoS$_2$ third‑body particles from their generation to ejection, demonstrating that the friction coefficient evolves through distinct time‑ or sliding‑distance‑dependent stages as the layer forms and stabilizes, ultimately decreasing sharply once a cohesive third‑body film is established.
The interactions of these particles within the contact zone are therefore crucial in determining overall tribological performance.

Wear particles can adopt two principal states: they may be active, meaning that they are capable of bearing part of the applied load, or inactive, when they become trapped in valleys or sintered into the surface and thus cease to participate dynamically in the tribological process \cite{Godfrey1954,Brink2021}.
The existence of these behaviors has been confirmed experimentally. In fretting tests, Colombié et al. \cite{Colombie1984} observed that wear debris formed a compacted third-body layer capable of carrying load and reducing friction, demonstrating the third body’s capacity to act as a solid lubricant while modifying local contact conditions. Similarly, Mischler and Martin \cite{Mischler2013} showed that, in metal-on-metal hip-joint simulators, wear particles can generate tribofilms that function both as protective and lubricating layers, reducing wear rates and improving tribological performance.
Godfrey and Bailey \cite{Godfrey1954} described how, in the early stages of fretting, some particles remained mobile while others became embedded or immobilized in surface asperities, indicating a transition between active and inactive states. 
Thus, when wear debris remain in an active state, they are capable of supporting load and acting as a lubricant by preventing direct metal-to-metal contact.

To understand how the contact state evolves in the presence of particles, one must examine particle–surface interactions at smaller scales. Discrete-element-method (DEM) studies have shown how particle cohesion, size, and shape govern transitions between fluid-like and solid-like behavior, thereby modifying effective friction \cite{Fillot2007, Basseville2011,Bilz2021,PhamBa2023}.
Hybrid FEM–DEM schemes now couple discrete particles with continuum deformation to capture both continuous and discontinuous material responses more efficiently. Voisin-Leprince et al.\ applied FEM–DEM coupling to simulate localized damage and wear phenomena at reduced computational cost \cite{Voisin-Leprince2024}. Similarly, Ahmadi and Sadeghi demonstrated how debris particles can evolve into protective platelets that significantly alter wear behavior \cite{Ahmadi2021}.
Studies by Wang et al. \cite{Wang2014} and Leonard et al.\ \cite{Leonard2014} confirmed experimentally observed third-body effects using FEM–DEM, highlighting the roles of particle agglomeration and cohesion in friction modulation. Nevertheless, DEM remains computationally intensive, particularly when resolving fine contact details.
Because friction is fundamentally an interfacial phenomenon, the boundary-element method (BEM) offers an attractive alternative to fully discretized or coupled models.

Lorig et al. first demonstrated that a two-dimensional DEM ring of jointed rock can be embedded within an outer elastic BEM halo, yielding a lightweight, reflection-free model for tunnel excavation \cite{Lorig1986}.
Huang et al. applied the same DEM--BEM coupling to simulate the penetration of a $60^{\circ}$ cone through sand grains, thereby capturing grain-scale force-chain mechanics \cite{Huang1993}.
Mirzayee et al. extended the concept to fluid--structure interaction by modeling a cracked concrete gravity dam with DEM blocks, while solving the surrounding infinite reservoir using BEM potential theory in a two-dimensional problem \cite{Mirzayee2011}.
Barros et al. presented a fully dynamic, monolithic two-dimensional coupling in which fractures in the DEM region generate stress waves that the time-domain BEM propagates to infinity. They later generalized the formulation to three dimensions for spherical particles \cite{Barros2023}.
Although the computational advantages of BEM are well known, its direct coupling with DEM remains rare in tribology; integrating the two therefore offers a promising compromise between computational efficiency and detailed particle dynamics.


In this work, we develop a simple numerical model for friction that accounts for the influence of wear debris during flat strip-draw friction tests. We begin with a simplified one-dimensional macroscale finite element (FEM) representation of the experimental setup. The effect of third-body particles is then incorporated within a local friction coefficient that depends on the area covered by active wear particles, serving as the state variable for the friction model. The local friction model is developed from a mesoscale numerical scheme that captures the interaction between third-body particles and the rough contacting surfaces. To achieve this, we introduce a novel simulation scheme that couples a rough-contact boundary element method (BEM) solver with discrete element method (DEM) particles representing the wear debris. The outputs of this mesoscale analysis are integrated into the macroscale model through homogenization. This multiscale coupling enables direct experimental validation and reveals how detailed mesoscale phenomena govern practical macroscale frictional behavior, thereby advancing both fundamental understanding and tribological modeling.

The paper begins with a description of the flat strip-draw test and the morphological analysis of debris, which motivates the development of the numerical model. The macroscale numerical model is then introduced, along with the governing equation for the local friction coefficient. Finally, the results of both mesoscale and multiscale simulations are discussed, and the paper concludes with a summary of the main findings.


\begin{center}
	\begin{longtable}{lll}
		\caption{Table of symbols used throughout the paper} \label{tab:symbols}                                                              \\
		\toprule
		\textbf{Symbol}                      & \textbf{Description}                                                           & \textbf{Unit} \\
		\midrule
		\endfirsthead

		\multicolumn{3}{c}%
		{{\bfseries Table \thetable\ Continued from previous page}}                                                                           \\
		\toprule
		\textbf{Symbol}                      & \textbf{Description}                                                           & \textbf{Unit} \\
		\midrule
		\endhead

		\bottomrule
		\multicolumn{3}{r}{{Continued on next page}}                                                                                          \\
		\endfoot

		\bottomrule
		\endlastfoot
		$F_N$                                & Normal force (macroscale)                                                      & N             \\
        $P_N$                                & Normal pressure (macroscale)                                                      & Pa             \\
		$F_S$                                & Shear force (macroscale)                                                       & N             \\
$h_B$   & Vertical thickness of of the deformable bar (macroscale)  & m \\
		$V_0$                                & Prescribed sliding velocity (macroscale)                                       & m/s           \\

		$v_0$                                & Prescribed sliding velocity (mesoscale)                                        & m/s           \\
		$\mu$                                & Local friction coefficient                                                     & --            \\
		$\bar{\mu}$                          & Average (macroscopic) friction coefficient,
		                                     & --                                                                                             \\
		                                     & defined as the ratio $\frac{F_S}{F_N}$ during the steady-state sliding regime. &               \\
        $ \sigma_{xx} $                                & Internal stress field in the bar   (macroscale)                                                             & Pa             \\
        $\Gamma_c$  & Contact region at the interface   (macroscale) & m \\
		$T_f$                                & Frictional traction   (macroscale)                                                             & Pa             \\
		$P$                                  & Normal pressure (macroscale)                                                   & Pa           \\
		$p$                                  & Normal pressure (mesoscale)                                                    & Pa           \\
		$T_f^{\max}$                         & Maximum admissible frictional traction                                              & Pa             \\
		$U$                                  & Tangential displacement (macroscale)                                           & m             \\
		$U^P$                                & Residual tangential displacement (macroscale)                                     & m             \\
		$k^s$                                & Shear stiffness                                                                & N/m           \\
		$\tau_0$                             & Shear strength of the softer material (aluminum)                               & Pa           \\
		$E^\star$                                & Effective elastic modulus                                                      & Pa           \\
		$\rho$                               & Mass density                                                                   & kg/m$^3$      \\
		$\beta$                              & Area fraction of third-body particles                                          & --            \\
		$K$                                  & Archard wear coefficient \cite{Archard1953}                                    & --            \\
		$H$                                  & Hardness of the softer material                                                & Pa           \\
		$j$                                  & Particle flux                                                                  & m/s           \\
		$\zeta$                              & Fraction of particles adhering to tool                                         & --            \\
		$\Delta t$                           & Time increment (macroscale)                                                    & s             \\
		$\Delta x$                           & Sliding distance (macroscale)                                                  & m             \\

		$dt$                                 & Time increment (mesoscale)                                                     & s             \\
		$\delta x$                           & Sliding distance (mesoscale)                                                   & m             \\
        
    $u$ & Normal displacement in the BEM, comprising elastic ($u_e$) and plastic ($u_p$) components & m \\
		$A_0$                                & Nominal contact area (mesoscale)                                                            & m$^2$         \\
		$A_c$                                & Real contact area (mesoscale)                                                              & m$^2$         \\
		$A_p$                                & Contact area of active  particles (mesoscale)                                            & m$^2$         \\
		$N_p$                                & Number of particles (mesoscale)                                                          & --            \\
		$N_c$                                & Number of contact clusters (mesoscale)                                                  & --            \\
		$N_s$                              & Number of surface summits (mesoscale)                                                        & --            \\
        		$\bar{d}$                            & Average spacing between summits (mesoscale)                                    & m             \\
		$d_c$                                & Distance between contact clusters (mesoscale)                                               & m             \\
		$\bar{h_g}$                                & Average gap height (mesoscale)                                                              & m             \\
		$d_p$                                & Particle diameter                                                              & m             \\
		$a_p$                                & Particle cross-section area                                                    & m$^2$         \\
		$h_p$                                & Particle height                                                                & m             \\
$\alpha$ & Parameter in the local friction coefficient model, dependent on the sliding velocity & -- \\

		$g(x,y)$                             & Gap profile between rough surfaces (mesoscale)                                 & m             \\
		$h_{\mathrm{Tool}}$                  & Height profile of rigid surface (mesoscale)                                    & m             \\
        		$h_{\mathrm{Tool}}^*$                & Modified tool surface height (mesoscale)                                       & m             \\
        		$h_{\mathrm{Def}}^0$                   & Initial height profile of deformable surface (mesoscale)                               & m             \\
		$h_{\mathrm{Def}}$                   & Height profile of deformable surface (mesoscale)                               & m             \\
$h_{\mathrm{Def}}^p$ & Plastically deformed profile of the deformable surface, defined as $h_{\mathrm{Def}}^0 - u_p$ (mesoscale) & m \\

$h'_{\mathrm{RMS}}$ & Root-mean-square (RMS) slope of the surface & -- \\

		$t_c(u)$                               & Contact traction in BEM                                                         & Pa           \\
		
		$z_p(x,y)$                           & Surface height of a particle                                                   & m             \\
		$\mathbf{r}$                         & Position of particle                                                     & m             \\
        		$m$                                  & Mass of particle                                                               & kg            \\
		$\mathbf{v}$                         & Linear velocity of particle (DEM)                                              & m/s           \\
		$\boldsymbol{\omega}$                & Angular velocity of particle (DEM)                                             & rad/s         \\

		$\mathbf{I}$                         & Moment of inertia of particle (DEM)                                            & kg·m$^2$      \\
		$\mathbf{f}^{\text{DEM}}$            & Forces from particle–particle interactions (DEM)                               & N             \\
		$\mathbf{f}^{\text{surface}}$        & Forces on a particle from interactions with rough surfaces (DEM)               & N             \\
		$\boldsymbol{\tau}^{\text{DEM}}$     & Torques from particle–particle interactions (DEM)                              & N·m           \\
		$\boldsymbol{\tau}^{\text{surface}}$ & Torques from from interactions with rough surfaces (DEM)                       & N·m           \\
		$\eta$                               & Damping coefficient in DEM                                                     & kg/s          \\

		$f(r)$                               & PDF for contact clusters (mesoscale)                                           & --            \\
		$\gamma$                             & Density in Poisson distribution (mesoscale)                                    & 1/m           \\
		$M_{IJ}$                             & Mass matrix in FEM                                                             & kg            \\
		$K_{IJ}$                             & Stiffness matrix in FEM                                                        & N/m           \\
		$T_{fI}$                             & Nodal force vector in FEM                                                            & N             \\
		$N_I(x)$                             & Shape function in FEM                                                          & --            \\
		$U_I$                                & Nodal displacement in FEM                                                      & m             \\
		$\dot{U}_I$                          & Nodal velocity in FEM                                                          & m/s           \\
		$\ddot{U}_I$                         & Nodal acceleration in FEM                                                      & m/s$^2$       \\
	\end{longtable}
\end{center}

\section{Experimental investigation}

\subsection{Flat strip-draw test setup}
\label{sec:stip-draw}
\begin{figure}[H]
\centering
	a \includegraphics[width=0.4\textwidth]{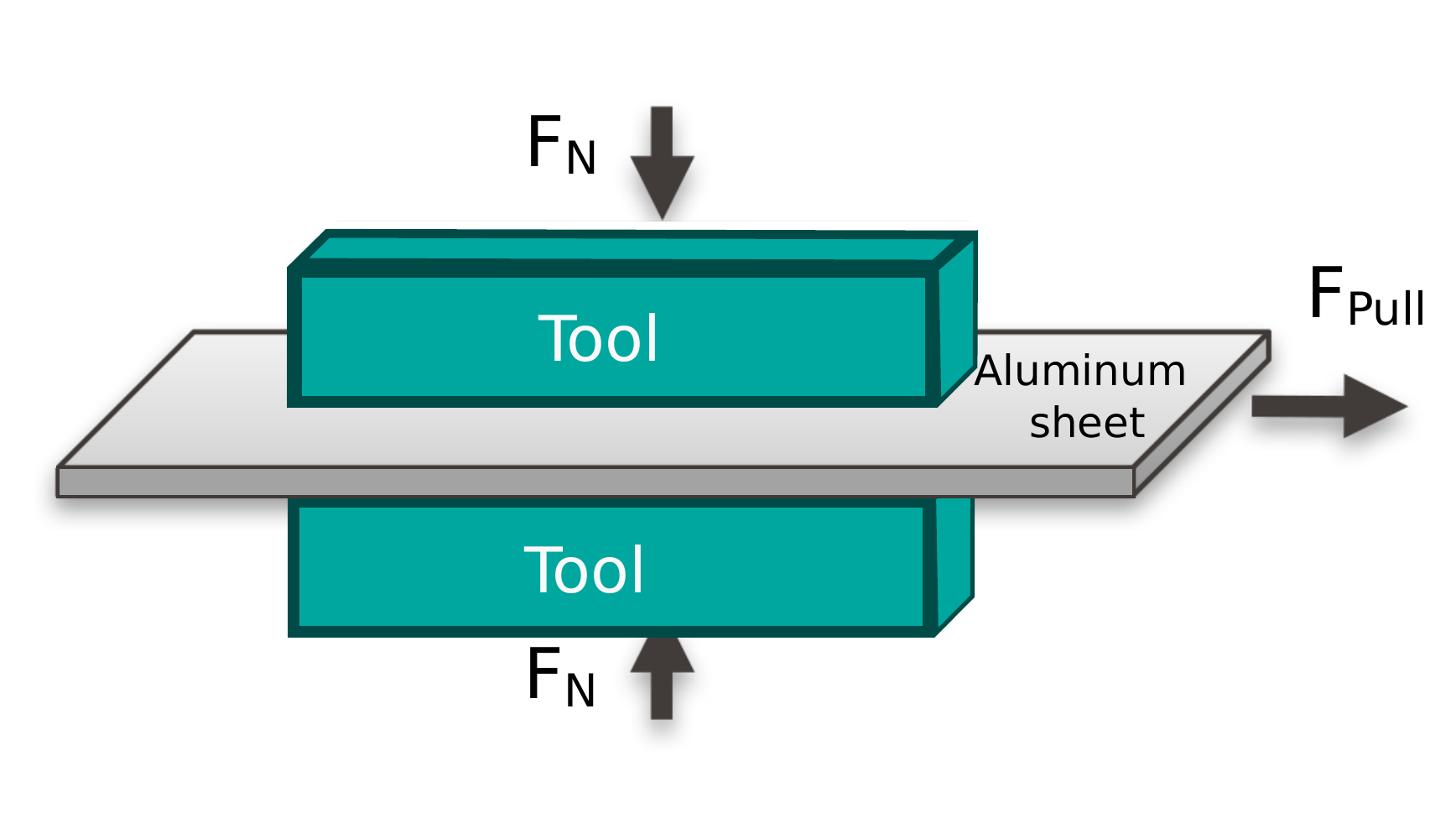}
    
    b \includegraphics[width=0.4\textwidth]{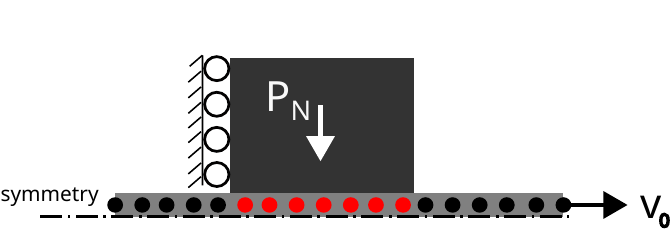}
	\caption{(a) Schematic of the flat strip-draw experimental test, where the tools are subjected to a normal load, and the aluminum sheet is pulled horizontally with a force $F_{\text{Pull}}$ at a prescribed velocity $V_0$.  
(b) Schematic of the macroscale simulation setup representing the experiment, in which a rigid tool is subjected to a normal pressure $P_N$ (resulting from the normal load $F_N$) and applied to an elastic one-dimensional bar that is pulled at a prescribed velocity $V_0$. The red points represent nodes on the bar that are in contact, while the black points indicate nodes that are either not yet in contact or have moved out of the contact zone. The number of nodes in the actual simulation is significantly higher than shown in the schematic. Due to geometric symmetry, only half of the domain is modeled in the numerical simulations.}
	\label{fig:exp}
\end{figure}

The experimental setup for the flat strip-draw friction test is illustrated in Figure \ref{fig:exp}(a). 
Aluminum sheets were placed between two flat cast-iron tool surfaces (EN-JS2070), which applied a constant normal load via load cells. The sheets were then pulled at a constant velocity of $0.08 \text{ m/s}$ along the length of the pads.

The tests were conducted using tools of varying lengths (35 mm, 70 mm, and 140 mm), while the tool width was kept constant at 35 mm. The aluminum sheet specimens were made of AA5182 alloy in the O-temper condition and featured an EDT (electro-discharge texturing) surface finish, with an arithmetical mean height of approximately $1,\text{\textmu m}$.
Each aluminum sheet had a width of 100 mm and a thickness of 1.5 mm. Both surfaces of the sheets were lubricated with a mineral oil-based lubricant, applied using electrostatic equipment, at a density of $1.12$ g/m$^2$.
Three tests were carried out to collect the final friction data under the same conditions to ensure reproducibility of results. During each test, the pulling force $F_{pull}$, the applied normal force, $F_N$, and the pulling displacement were continuously recorded. The time-dependent coefficient of friction was calculated as:
\begin{equation}
	\mu(t) = \frac{F_{pull}(t)}{2F_N(t)}
	\label{eq:mu_exp}
\end{equation}

\begin{figure}[H]
 \centering
   a \includegraphics[width=0.5\textwidth]{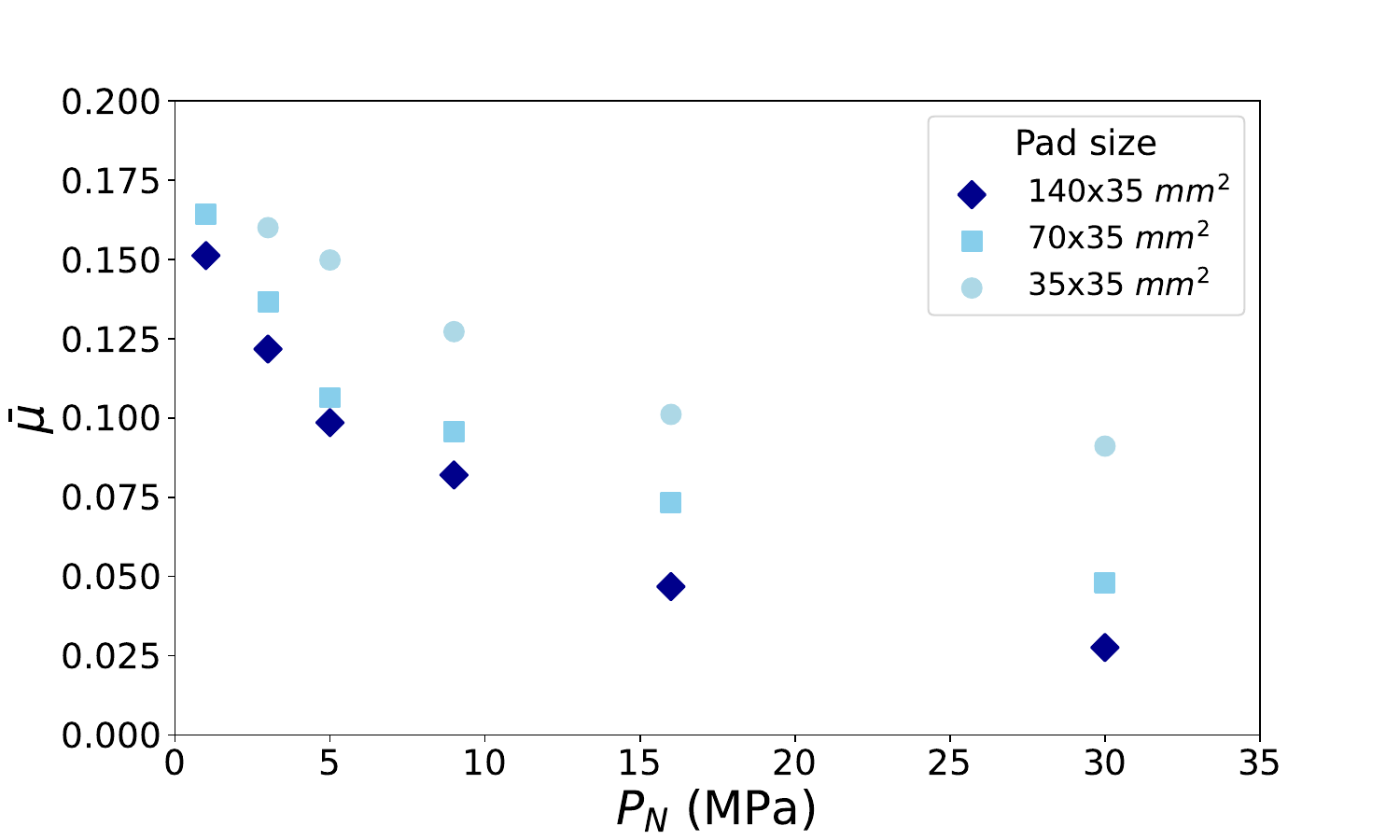}
  
    b\includegraphics[width=0.5\textwidth]{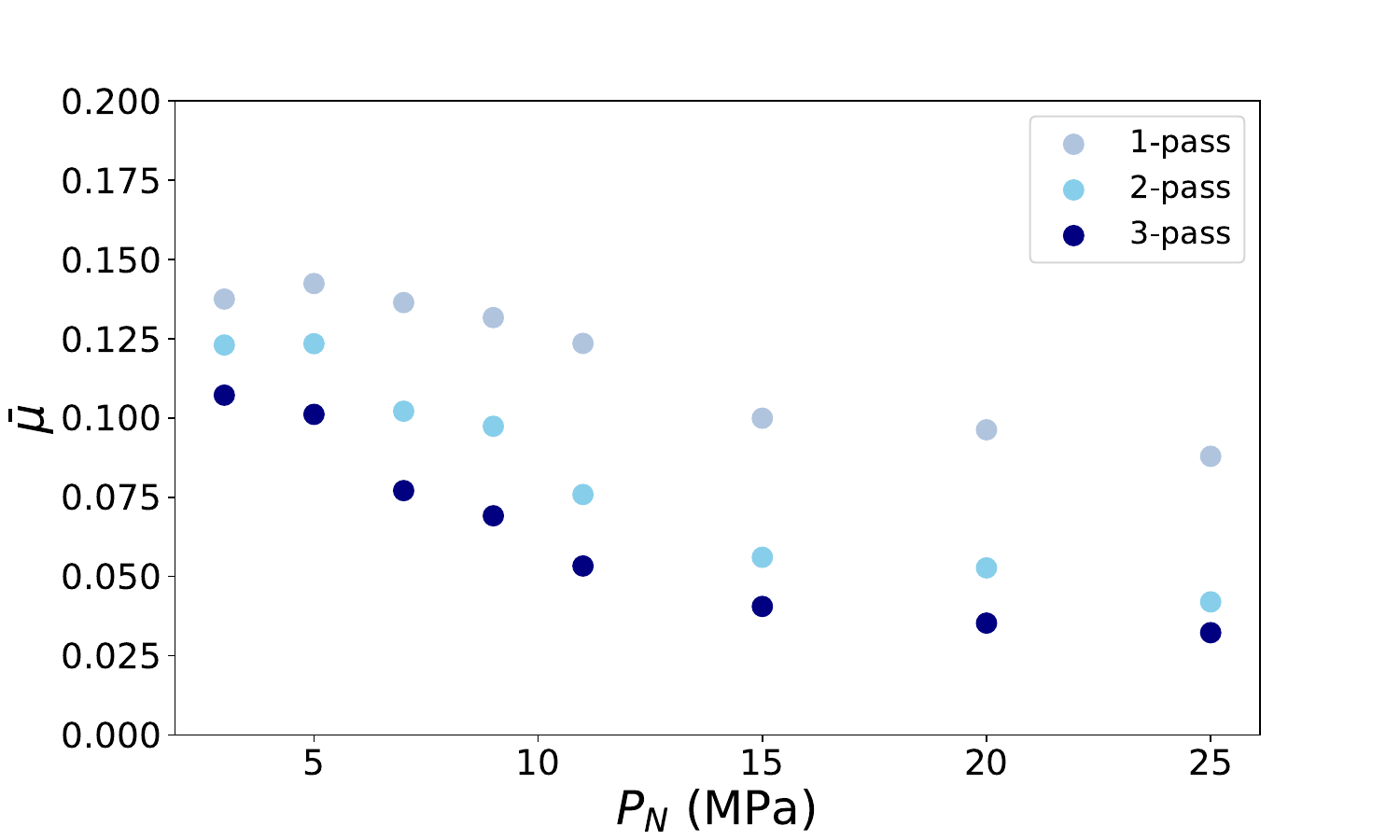}
   
  \caption{Evolution of the average friction coefficient, denoted $\bar{\mu}$, over the converged portion of $\mu(t)$, as a function of applied normal pressure $P_N$:  
(a) for different tool-pad lengths (the influence of tool-pad width was minimal and is shown in Appendix~\ref{sec:pad-width});  
(b) for multi-pass strip-draw tests using a tool 35~mm long and 35~mm wide.}

  \label{fig:combined_experiments}
\end{figure}

The measured friction coefficients during each test initially exhibited a peak, followed by a gradual decrease, and eventually converged to a steady-state value. The average coefficient of friction reported in this study was extracted from the steady-state sliding region near the end of each test, where the frictional response had stabilized.
Figure \ref{fig:combined_experiments}(a) shows the average coefficient of friction, $\bar{\mu}$, as a function of the applied normal pressure, $P_N$, for three different tool‐pad sizes. 
The test results indicate that the average friction coefficient decreases significantly with increasing normal pressure. 
Moreover, it is evident that the pad size has a noticeable effect on $\bar{\mu}$.

Repeating the test without cleaning the surface (2-pass and 3-pass tests in Figure \ref{fig:combined_experiments}(b)) results in an even greater reduction in the average friction coefficient with increasing pressure, up to 72\% at 25 MPa, compared to a 35\% reduction after the first pass.

In contrast, tests in which the surface was cleaned, re-lubricated, and then retested show little difference compared to the 1-pass sample.

One of the main observations during the strip-draw tests was the formation of a grayish dust on the surface after testing, consisting of small wear particles generated during sliding.
These findings suggest a correlation between the amount of wear debris and the reduction in friction coefficient on aluminum surfaces. This observation motivates the development of a friction model capable of predicting the macroscopic friction coefficient while accounting for the presence of third-body particles in the contact interface.

\subsection{Morphological analysis of wear debris}
\begin{figure}[H]
	\centering
	\includegraphics[width=0.4\textwidth]{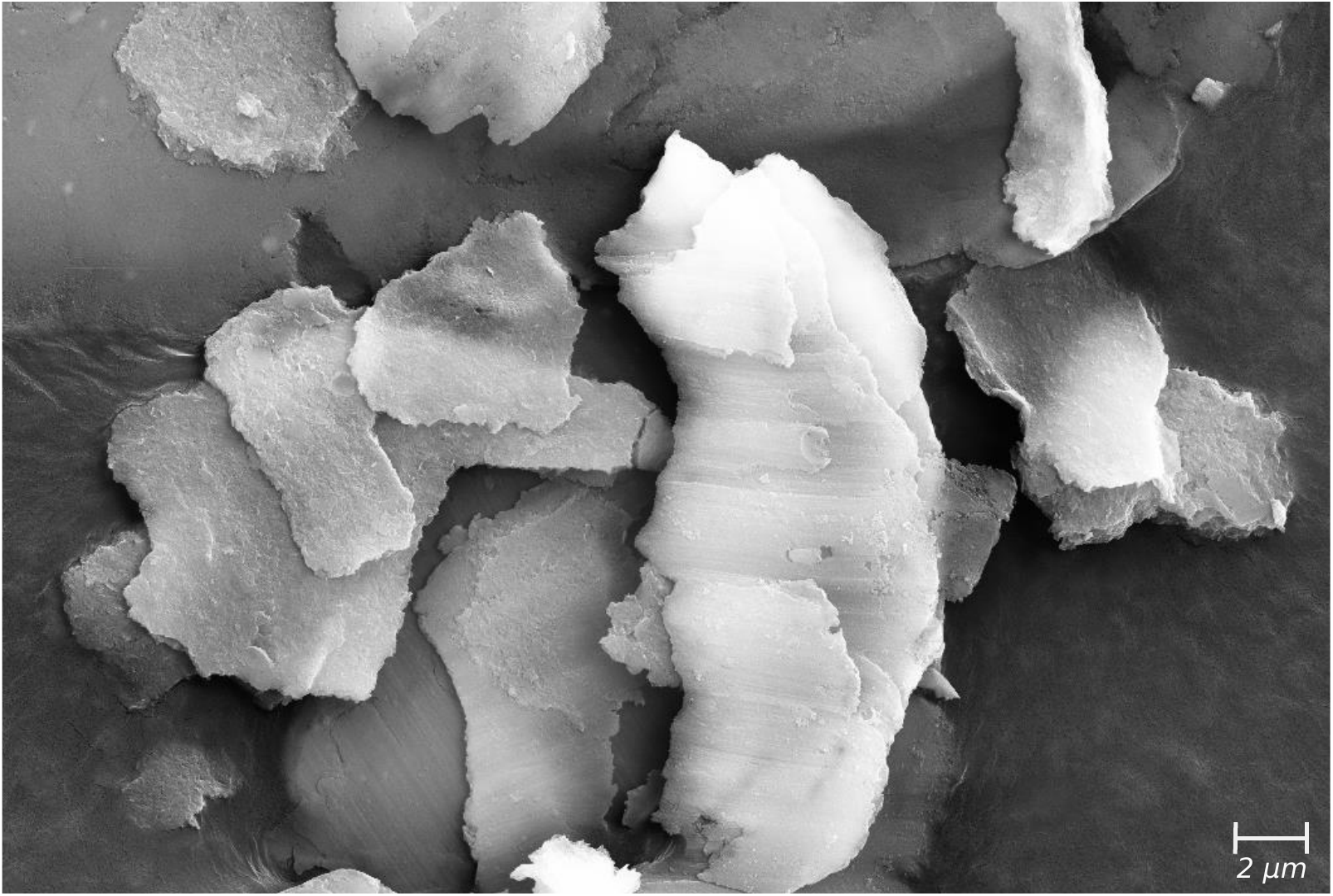}
	\caption{
		SEM image of wear debris generated during the strip-draw friction test. The particles exhibit a flake-like morphology, characterized by large lateral dimensions relative to their thickness.
	}

	\label{fig:SEM_image}
\end{figure}

The analysis of wear particle morphology created during the strip-sraw friction test of aluminum samples was performed using scanning electron microscopy (SEM). The SEM image in Figure \ref{fig:SEM_image} showed that the particles were flake-like, exhibiting lateral dimensions much larger than their thickness. To further investigate the composition of the tribolayer and adhered particles, energy-dispersive X-ray spectroscopy (EDS) was carried out on three points across the surface.
The elemental composition (obtained from three different locations) is presented in Table \ref{tab:EDS_composition}. The detected carbon originates from the carbon tape used during the test as a conductive support. The results show that the particles consist mainly of aluminum and aluminum oxide. 

\begin{table}[h!]
	\centering
	\caption{Elemental composition of the tribolayer from SEM-EDS (in weight\%)}
	\label{tab:EDS_composition}
	\begin{tabular}{lcccccccc}
		\toprule
		Element    & C     & O     & Mg   & Al    & Si   & Ti   & Mn   & Fe   \\
		\midrule
		Spectrum 1 & 13.63 & 1.30  & 2.74 & 82.06 & 0.00 & 0.05 & 0.22 & 0.00 \\
		\midrule
		Spectrum 2 & 19.68 & 20.26 & 2.48 & 57.27 & 0.00 & 0.19 & 0.13 & 0.00 \\
		\midrule
		Spectrum 3 & 24.72 & 17.13 & 2.14 & 55.58 & 0.00 & 0.22 & 0.21 & 0.00 \\
		\bottomrule
	\end{tabular}
\end{table}

\subsection{Particle size distribution analysis}

\label{sec:setup-meso-particle-size}
\begin{figure}[H]
	\centering
	\includegraphics[width=0.55\textwidth]{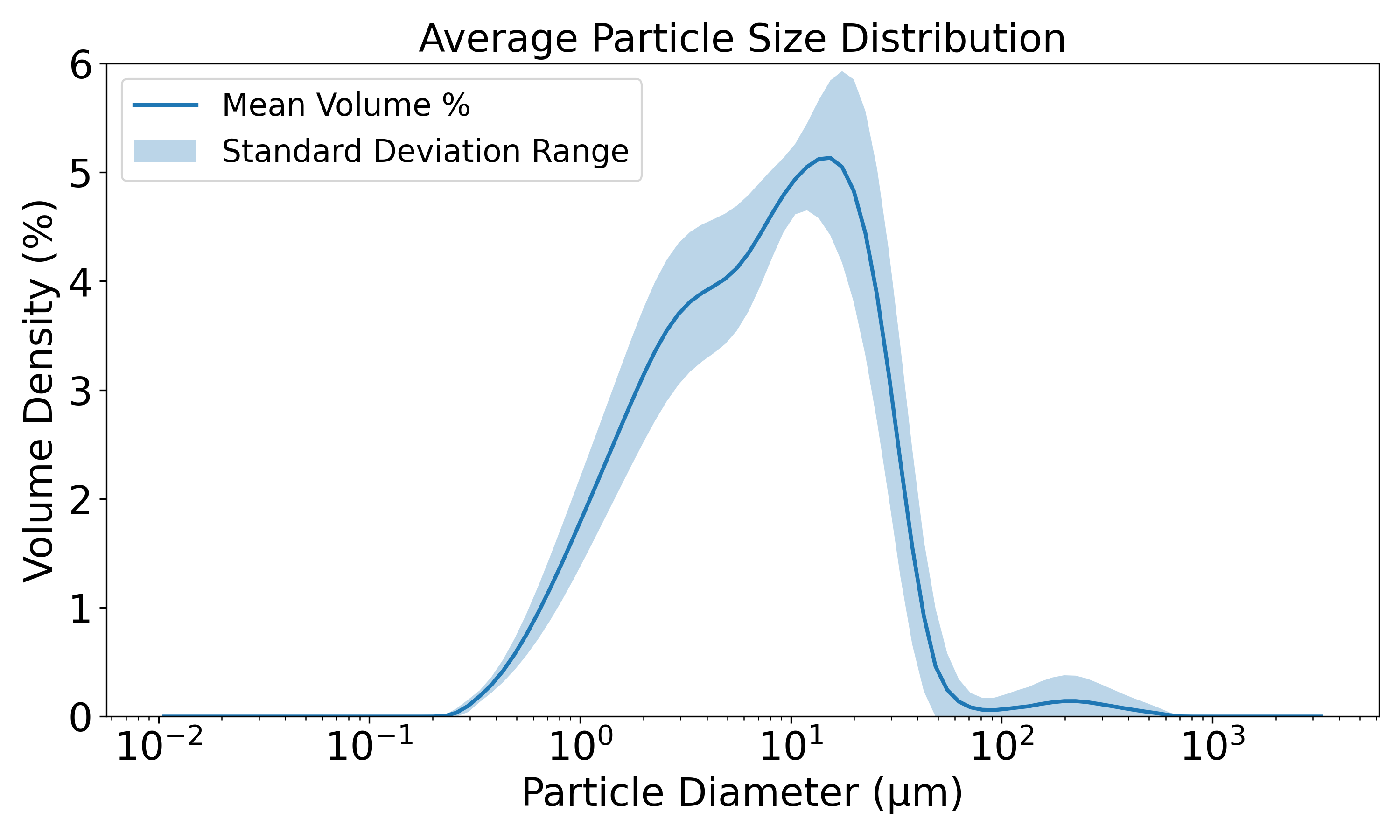}
	\caption{Average particle size distribution of flake-like particles, measured by laser diffraction (Mastersizer 3000, Malvern Panalytical) in toluene dispersion mode. The curve shows the mean volume percentage as a function of particle diameter, expressed as the volume-equivalent spherical diameter. The shaded region represents the variation across all measured samples under applied normal pressures from 1 to 13 MPa (±1 standard deviation). Each test was repeated twice to confirm reproducibility. No clear trend indicating a dependence of particle size on applied normal pressure was observed. The x-axis is shown on a logarithmic scale to capture the full size range.}
	\label{fig:particle-size}
\end{figure}

The particle size distribution was measured using a Mastersizer 3000 laser diffraction system (Malvern Panalytical). Measurements were conducted in wet-dispersion mode, using toluene as the medium. As laser diffraction assumes spherical particles, the results are reported as volume-equivalent spherical diameters. Since the particles possess a flake-like geometry, the measured diameters primarily represent the lateral size and not the actual thickness. Therefore, particle sizing results should be interpreted in conjunction with SEM images for accurate assessment of particle dimensions. As shown in the SEM images in Figure \ref{fig:SEM_image}, the particles exhibit a flake-like morphology, and their thickness appears to be sub-micron. In the simulations we therefore assumed a particle height of $0.5 \text{  \textmu m}$. For the particle diameter, we considered several values between 6 and 100 $ \text{  \textmu m}$, which fall within the measured diameter distribution (see Figure \ref{fig:particle-size}).

\subsection{Surface roughness characterization via confocal microscopy}
\label{sec:setup-meso-confocal}
\begin{figure}[H]
	\centering

	a\includegraphics[width=0.35\textwidth]{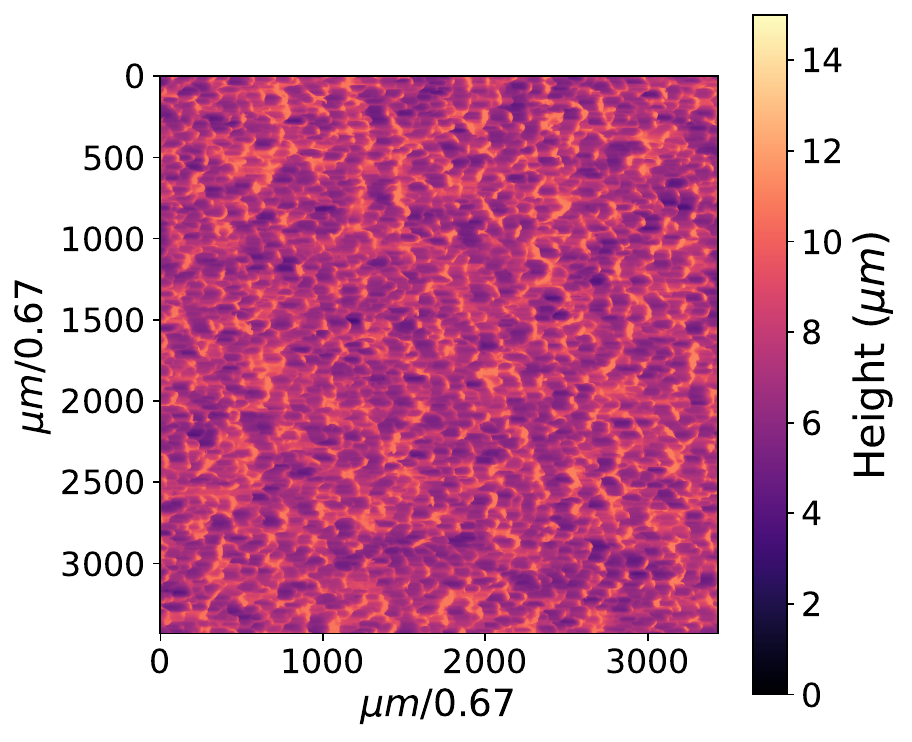}
	b\includegraphics[width=0.35\textwidth]{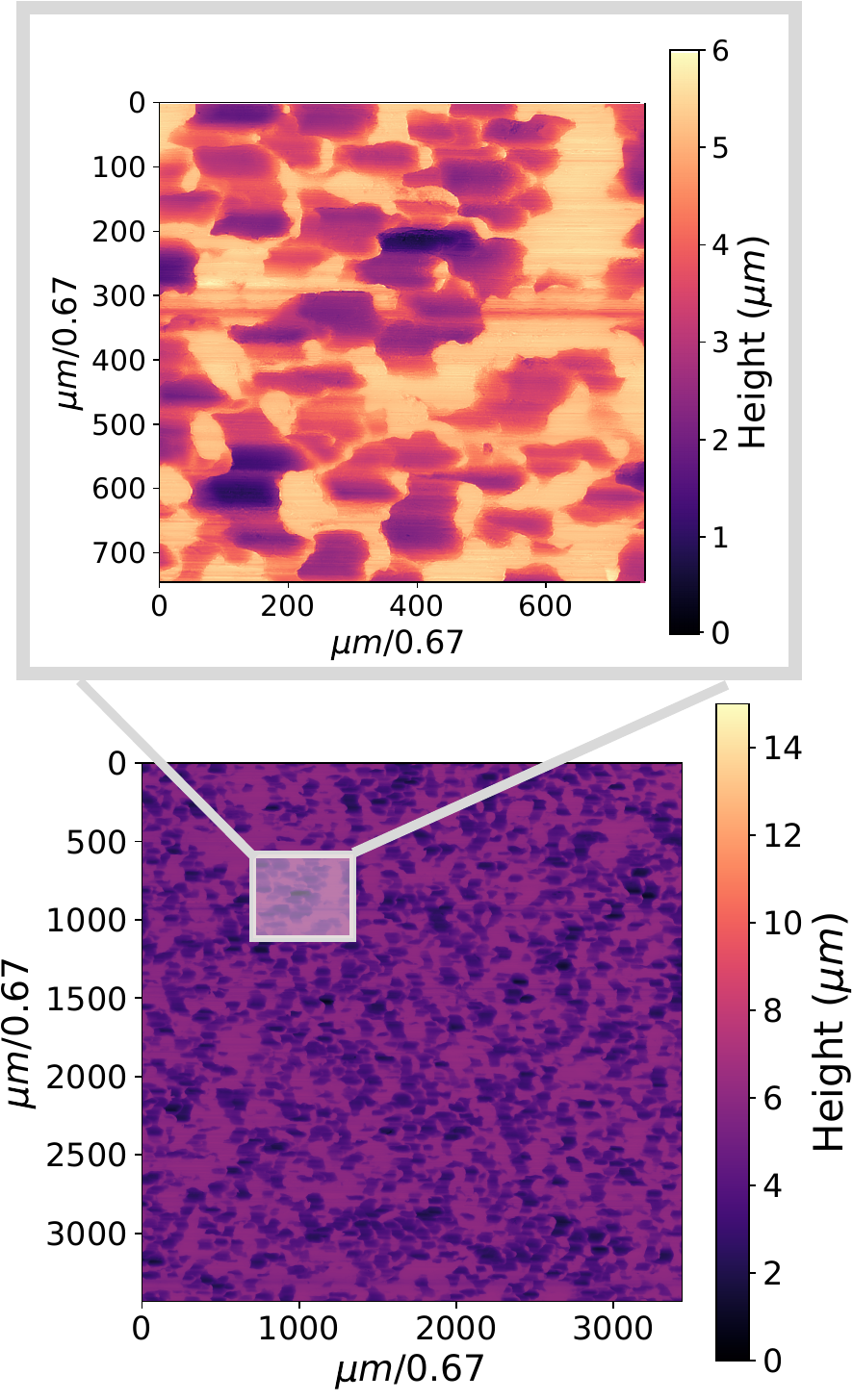}
	c\includegraphics[width=0.35\textwidth]{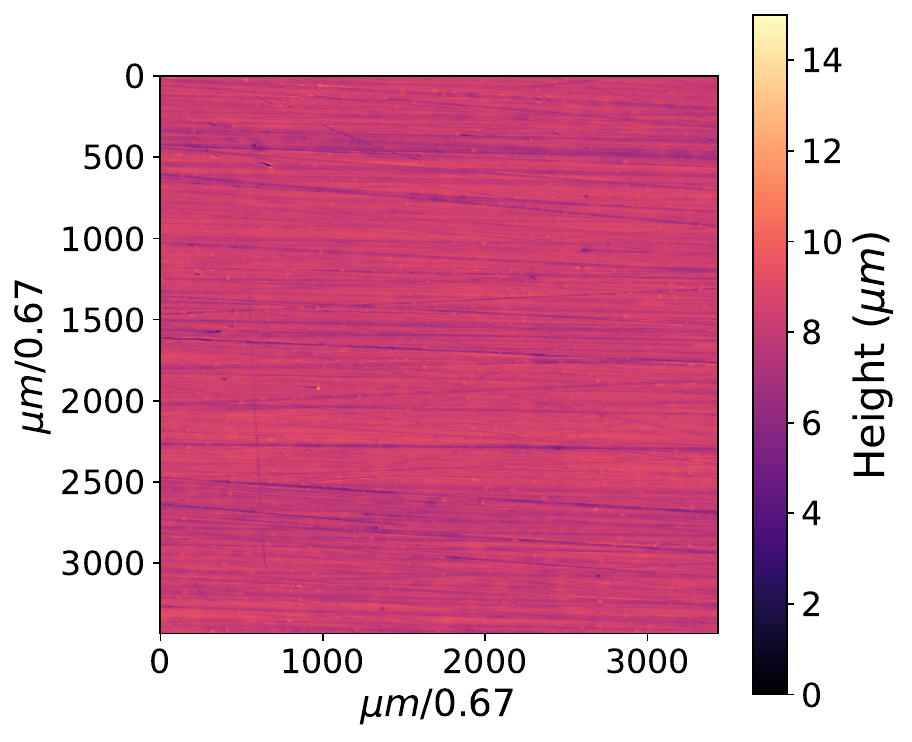}
	\caption{Surface topography measured by confocal microscopy. (a) Confocal image of the aluminum sheet surface before strip-draw testing. (b) Aluminum surface after the application of 10 MPa normal pressure and 90 mm of sliding, at the exact same location as (a). (c) Surface profile of the cast-iron tool used in the strip-draw friction test. 
	}
	\label{fig:confocal_PSD}
\end{figure}

Surface topography measurements of the aluminum samples and tool surfaces were carried out using a confocal microscope, focusing on regions far from the sample edges. This technique employs a focused laser beam and a pinhole aperture to eliminate out-of-focus light, enabling high-resolution, three-dimensional mapping of surface heights. 
Prior to lubrication and friction testing, each sample undergoes a standardized surface characterization procedure. A microhardness imprint of approximately 100 µm is first applied to the surface as a positional reference. This step is performed before any surface treatment to ensure an accurate baseline for imaging. Confocal microscopy at 20× magnification is then used to capture the surface texture. Each scanned area measures $2 \times 2$ mm$^2$ and is recorded with $3432 \times 3432$ points, corresponding to a lateral spacing of $0.67~\text{\textmu m}$ between adjacent nodes. Following imaging, the sample is subjected to oiling and strip-draw friction testing. After testing, the surface is cleaned with isopropanol, and a second confocal image is acquired at the exact location of the initial measurement by relocating the original hardness imprint. This approach ensures a consistent spatial comparison of surface features before and after friction testing.
The resulting measurements served as a guideline for generating the initial periodic self-affine surface roughness used in the simulation.
Figure~\ref{fig:confocal_PSD} presents the surface-height profiles measured before and after the strip-draw friction test conducted at a normal pressure of $10~\text{MPa}$ and a sliding distance of $90~\text{mm}$.

For the aluminum surface, the arithmetical mean height fell from approximately $1.16~\text{\textmu m}$ before the test to $0.81~\text{\textmu m}$ afterwards, while the root--mean--square (RMS) height decreased from $1.40~\text{\textmu m}$ to $0.95~\text{\textmu m}$. These reductions indicate modest surface \emph{smoothing}, most likely caused by asperity flattening and abrasion. 
In addition to this smoothing effect, horizontal lines appear across the aluminum surface after sliding, which are especially visible in the zoomed‑in region of Figure \ref{fig:confocal_PSD}(b).
The RMS slope remained nearly constant ($\approx 0.34$) before and after sliding, showing that the average surface gradient stayed low.

The cast--iron tool surface exhibited a much lower arithmetical mean height ($\approx 0.44~\text{\textmu m}$) and an RMS height of $\approx 0.60~\text{\textmu m}$. 

\begin{figure}[H]
	\centering
a\includegraphics[width=0.4\textwidth]{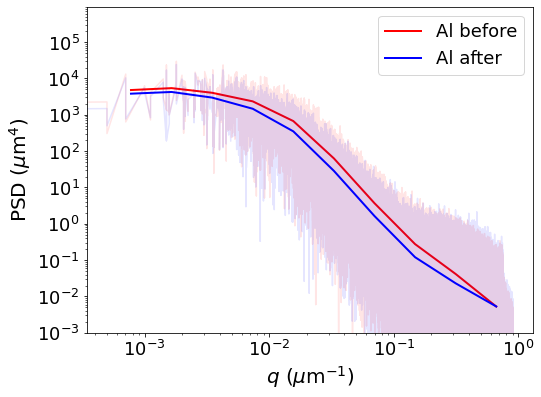}
b\includegraphics[width=0.4\textwidth]{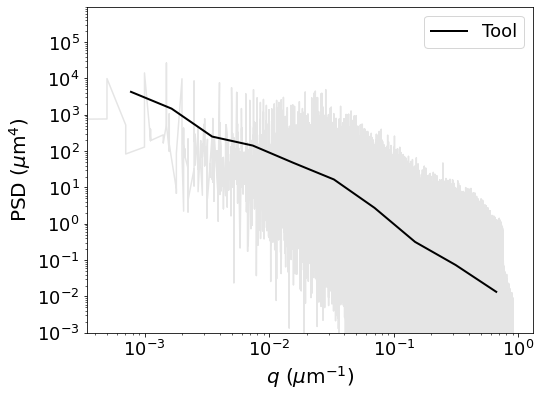}
	\caption{Power spectral density (PSD) profiles obtained from confocal surface measurements. (a) Aluminum surface before and after the strip-draw test, performed at the same location under a normal pressure of 10 MPa and a sliding distance of 90 mm. (b) Cast-iron tool surface. The solid lines represent the radially averaged PSD, while the shaded regions correspond to the raw, unaveraged PSD data. The averaged PSD profiles of the aluminum surface before testing and of the tool are used to generate the initial surface topography for the mesoscale simulations.}
	\label{fig:exp_PSD}
\end{figure}

The radially averaged power spectral density (PSD) was also extracted from the surface measurements, as shown in Figure \ref{fig:exp_PSD}. The PSD profile indicates that the average magnitude of the radially averaged PSD for aluminum decreases with sliding, suggesting that the surface becomes smoother during the strip‑draw friction test. The averaged PSD profiles of the aluminum surface before the test and the tool are used to generate the periodic self-affine surface employed in our simulations.

%

\section{Macroscale numerical model \label{sec:macro-num-model}}
As shown in Figure \ref{fig:exp}(b), the macroscale finite‑element (FE) model
consists of a rigid tool applying a normal load to a one‑dimensional,
deformable (elastic) bar. The bar translates horizontally
with an average sliding velocity $V_{0}$, and its right‑most point moves
exactly at $V_{0}$. This model provides a simplified representation of
macroscale experimental strip‑draw friction tests.
Let $U(x,t)$ denote the horizontal displacement at position $x$ and time $t$ along the bar. The governing equation for motion in the bulk of the bar is given by:
\begin{equation}
   \rho\, \ddot{U}(x,t)
  = E^{*}\, \nabla^{2}U(x,t),
  \label{eq:macro_u_corrected}
\end{equation}
where $\rho$ is the material density and $E^{*}$ is the effective Young’s modulus. The deformation is driven by contact interaction with a rigid tool acting on the lower surface of the bar. A Neumann-type boundary condition is imposed at the contact interface:
\begin{equation}
    \sigma_{xx}(x \in \Gamma_c, t) = T_f(x,t),
\end{equation}
where $\sigma_{xx}$ denotes the internal axial stress in the bar, $\Gamma_c$ is the contact region (highlighted as red nodes in Figure \ref{fig:exp}(b)), and $T_f(x,t)$ is the tangential traction (friction pressure) acting in the $x$‑direction at the contact interface.
The traction $T_{f}(x,t)$ is defined according to an elastic–perfectly‑plastic Coulomb law:
\begin{equation}
  T_{f}(x,t)=
  \begin{cases}
    k_{s}\,\bigl[U(x,t)-U^{\mathrm p}(x,t)\bigr],
      & \text{if }
        \bigl|k_{s}[U(x,t)-U^{\mathrm p}(x,t)]\bigr|
        \le T_{f}^{\max}(x,t), \\[6pt]
    \operatorname{sgn}\!\bigl(k_{s}[U(x,t)-U^{\mathrm p}(x,t)]\bigr)\,
      T_{f}^{\max}(x,t),
      & \text{otherwise},
  \end{cases}
  \label{eq:friction_law}
\end{equation}
with the local yield strength defined as
\begin{equation}
  T_{f}^{\max}(x,t) = P(x,t)\, \mu(x,t).
\end{equation}
Here, $k_s \propto E^{*} / h_B$ denotes the tangential (shear) penalty stiffness of the interface spring, where $h_B$ is the bar height, $U$ is the current tangential displacement, and $U^{\mathrm{p}}$ is the accumulated irreversible slip.
The normal pressure is denoted by $P(x,t)$, and the friction coefficient $\mu(x,t)$ evolves based on the state of third‑body particles, characterized by the internal variable $\beta(x,t)$.
The only remaining unknown is the evolution of $\mu$ in terms of the mesoscale state variable $\beta$ and the loading conditions. Equation \ref{eq:macro_u_corrected} is then discretized via FEM \cite{Logan1997} and integrated in time using the Newmark–$\beta$ scheme (see Appendix~\ref{sec:Macro-formulation}). It is worth noting that our friction model should not be confused with classical rate–state formulations, which account for typical rate effects (e.g., velocity weakening or strengthening), and can reproduce stick–slip motion through the evolution of a state variable. Here, the proposed model is a simple Coulomb friction law aimed at capturing the time-dependent influence of third-body particles on the macroscale frictional response.
%

\section{Governing equation for the local friction coefficient}\label{sec:coupling-macro-meso}
At each node of the macroscale FEM simulation, we need to define the local friction coefficient. Here, we use Bowden and Tabor friction model \cite{Bowden}, which assumes that the frictional force is proportional to the real contact area $A_c$:
\begin{equation}
	\mu_0=\frac{A_c \tau_0}{A_0 P},
\end{equation}
where $\tau_0$ is the shear strength of the material, $A_0$ is the nominal contact area and $P$ is the applied pressure.
To account for the influence of the third body on the friction coefficient, we assume that the particles are unbreakable and friction-less flakes. Therefore, particles trapped between rough surfaces can carry the normal contact load while reducing shear resistance, thereby acting as a lubricating layer. This is consistent with experimental observations (see Section \ref{sec:results} and references \cite{Diomidis2011,Ciprari2023}), which show that increasing particle density leads to a lower friction coefficient. As a result, the hypothesis that the shear force is reduced proportionally to the area covered by the particles leads to a modified friction coefficient:

\begin{equation}
	\mu = \frac{(A_c-A_p)\tau_0}{P A_0}=
    \frac{A_c \tau_0}{P A_0}\left(1 - \frac{A_p}{A_c}\right),
	\label{eq:mu}
\end{equation}
where $ A_c $ is the total real contact area, and $ A_p $ is the area where the normal load is carried by particles, i.e., the area where particles are activated as lubricant. The total real contact area, $ A_c $, can be computed using the BEM--DEM framework or estimated with plasticity-based models, such as those in \cite{Bush,molinari}.
\begin{equation}
	A_c = A_0\,\frac{\kappa}{h'_{\mathrm{RMS}}}\,\frac{P}{E^\star},
	\label{eq:Ac}
\end{equation}
where  $ E^\star $ is the effective elastic modulus, $ \kappa $ is a constant derived from plasticity theories, and $ h'_{\mathrm{RMS}} $ is the root-mean-square slope of the surface (the comparison between Equation \ref{eq:Ac} and BEM simulation can be found in Appendix \ref{sec:Ac-comparision}). This yields the relation:
\begin{equation}
	\mu = \tau_0\left(\frac{\kappa}{E^\star h'_{\mathrm{RMS}}} - \frac{A_p}{P A_0}\right),
	\label{eq:mu2}
\end{equation}
Making the quantity $A_p / A_0$ a critical factor to characterize 
particles' influence over friction.
Nevertheless, $A_p$ remains to be determined in this formulation (Equation \ref{eq:mu2}), and it necessarily depends on the applied normal pressure $P$, the sliding velocity $\dot{U}$, and the amount of third-body particles.
Given their flake-like geometry, the quantity of third-body particles in our model is characterized by the area fraction of particles covering the surface, as follows:
\begin{equation}
	\beta = \frac{a_p N_p}{A_0},
	\label{eq:beta}
\end{equation}
where $a_p$ is the average cross-sectional area of the particles, and $N_p$ is the number of particles.  
Here, we define the quantity of particles as the state variable $\beta$.  
Since the contact area covered by active particles, $A_p$, should not exceed the total real contact area, $A_c$, we introduce a threshold on $\beta$ in our formulation, denoted as $\beta^{\text{thres}}$, which may lead to saturation. Accordingly, $\beta$ is restricted to the range $0 < \beta < \beta^{\text{thres}}$.  
Finally, the area covered by active particles is defined as a function of pressure, sliding velocity, and particle quantity:
\begin{equation}
    A_p \equiv A_p(P, \dot{U}, \beta)
\end{equation}

The quantity of particles, $\beta$, evolves with time, proportionally with the sliding distance, as particles are created with wear.
Moreover, particles may be transported, i.e., staying with the surface and/or with the tool. The conservation of particles can be described with a transport equation:
\begin{equation}
	\frac{\partial \beta}{\partial t} + \nabla \cdot j = K \frac{P}{H} \frac{|\dot{U}|}{h_p},
	\label{eq:beta_evolution}
\end{equation}
where $K$ is the Archard wear coefficient, $h_p$ is the particle thickness, and $j$ is the particle advective transport rate with the tool.
The right hand side is the particle source term, modeling particles creation due to wear, following Archard's wear law \cite{Archard1953} (see Appendix \ref{sec:Archard} for details),
while $j$ is defined as:
\begin{equation}
	j = -\zeta \dot{U} \beta,
\end{equation}
with $\zeta$ representing the fraction of particles sticking to the tool. $\zeta \in [0,1]$, with $\zeta=1$ meaning all particles stay within tool asperities, whereas $\zeta=0$ meaning that all particles stay within asperities of the moving bar.

An upwind finite difference scheme is employed to compute,
from Equation \ref{eq:beta_evolution},
the evolution of $\beta$, at each node $i$:

\begin{equation}
	\beta_i^{n+1} = \beta_i^{n} + \Delta t \left[ K \frac{P_i}{H} \frac{|{\dot{U}}|}{h_p} + \frac{\zeta}{l_e}\left(  \dot{U}_{i+1} \beta_{i+1} -\dot{U}_i \beta_i \right) \right],
	\label{eq:beta_update}
\end{equation}
where $\Delta t$ is the time step increment, and $l_e$ is the mesh element size.
The explicit $l_e^{-1}$ factor in the upwind discretisation of the $\beta$‑transport equation introduces a mild mesh dependence; however, an $h$‑refinement study performed by halving the element size confirmed that its effect on the final $\beta$ field remains below 1\%.
Moreover, dynamic effects caused by inertia (i.e., the left-hand side term in Equation \ref{eq:macro_u_corrected}) can result in a faster, more localized, or more oscillatory evolution of $\beta$ by influencing the sliding velocity, $\dot{U}$. The updated particle fraction $ \beta $ is then used to compute $ A_p $ on each node in contact and, consequently, the local friction coefficient $ \mu $.

The relationships embedded in $A_p(P, \dot{U}, \beta)$ reflect the propensity of particles to lubricate the contact: it highly depends on the topography and deformation of contacting surfaces. To determine this relationship, we have developed a mesoscale numerical model, which is described in the following section.

\section{Mesoscale numerical model}
\label{sec:coupling-bem-dem}
This section presents a coupled numerical approach to model the interactions between two rough surfaces in the presence of particles within the separation gap. 
The model aims to mimic what occurs at any node of the macroscale simulation.
The rigid upper surface (the “tool”) is subjected to a constant normal pressure $p$ (corresponding to the nodal pressure $P$ in the macroscale model) and a constant sliding velocity $v_0$ (corresponding to the nodal velocity $\dot U$ in the macroscale model).
We assume that $v_0$ is much lower than the elastic wave‐propagation speed in the bulk material and that any particles trapped in the gap behave as perfectly rigid. The lower surface (the "deformable") is considered elasto-plastic as described further below.

\begin{figure}[H]
	\centering
	\includegraphics[width=\linewidth]{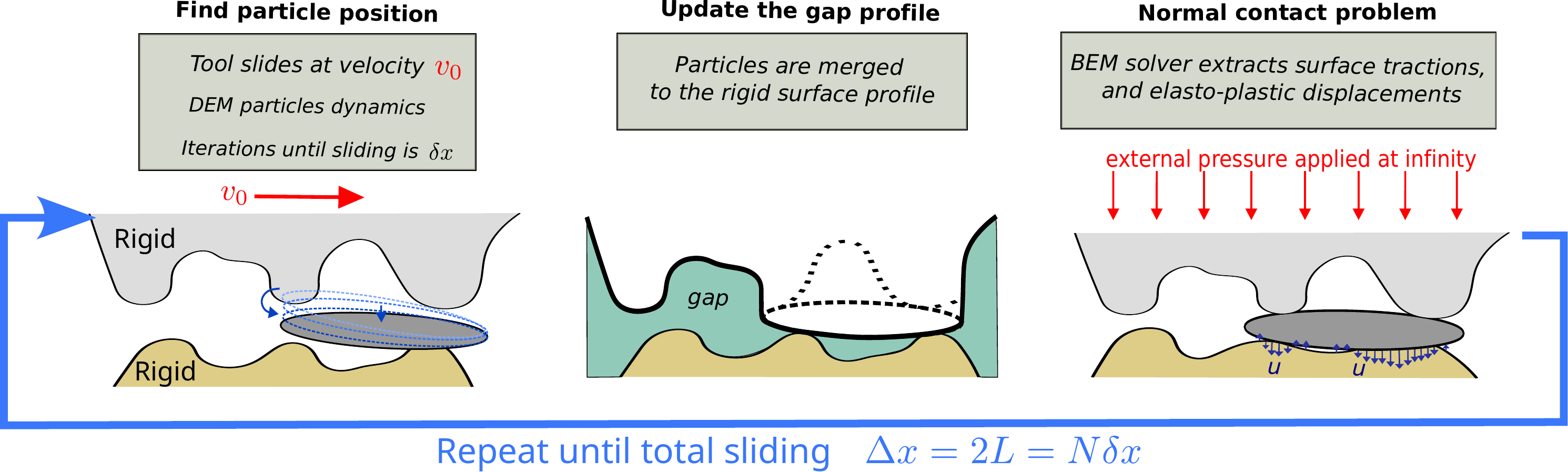}
    
	\caption{A schematic illustration of the coupled BEM–DEM code. To find the particles’ positions, rigid sliding is conducted during a DEM simulation at a velocity of $v_0$ until the tool advances by one pixel, $\delta x$. Then, a BEM solution allows to extract the deformation of the lower surface, assuming the particles are rigidly attached to the tool. These steps are repeated until the tool reaches the desired total sliding distance.}
	\label{fig:schem_coupled}
\end{figure}

Both surfaces are discretized on a uniform grid with spacing $\delta x$, as is standard when using a Fourier–BEM solver \cite{PolonskyKeer1999}.
As a result, the gap profile $g$, which is classically used in the BEM contact solver, can only be computed when the top surface slides by an integer multiple of the grid resolution $\delta x$ (i.e., one pixel).
Therefore, the staggered scheme illustrated in Figure \ref{fig:schem_coupled} is employed. First, the positions of the particles are computed while the rough surfaces remain rigid, and the top surface (tool) slides until a distance of one pixel is reached. Blocked particles are then localized (see details below). Subsequently, the elasto-plastic normal contact problem is solved using modified surface roughnesses (i.e., gap profiles) that account for the presence of these blocked particles.


%
These stages are repeated until the total sliding distance $\Delta x = 2L$ is reached, which is independent of $v_0$, as $L$ denotes the width of the periodic domain.

\textbf{Step 1 -- Finding particles' positions:} The  trajectories along a sliding distance $\delta x$ are established within $n$ dynamical time steps, where $n v_0 dt = \delta x$ ($dt$ is a timestep satisfying CFL stability conditions).  During each time step, particles' evolution is computed using the velocity-Verlet algorithm \cite{Verlet1967}, which updates the position and velocity of each particle from
the following translational and rotational equations of motion:
\begin{equation}
	m \ddot{\mathbf{r}} = \mathbf{f}^{(\mathrm{DEM})} + \mathbf{f}^{(\mathrm{surface})} - \eta \mathbf{v},
	\label{eq:dem-linear}
\end{equation}
\begin{equation}
	\mathbf{I} \dot{\boldsymbol{\omega}} = \boldsymbol{\tau}^{(\mathrm{DEM})} + \boldsymbol{\tau}^{(\mathrm{surface})} - \eta \boldsymbol{\omega}.
	\label{eq:dem-rotation}
\end{equation}
Here, $ \mathbf{r}$ is the position of the particle, $ \mathbf{v}$ its linear velocity, $ m $ its mass, $ \boldsymbol{\omega} $ its angular velocity, and $ \mathbf{I} $ its moment of inertia.
In this model, two types of contact interactions are considered, both issued from a penalty method. The penalty stiffness was chosen so that a maximum overlap of 5\% of particles' thickness, while the maximum contact force was estimated using Hertzian contact theory \cite{Johnson1985}.
$\mathbf{f}^{(\mathrm{DEM})}$ includes particle–particle contact forces, while $\mathbf{f}^{(\mathrm{surface})}$ accounts for particle–surface interaction forces arising from interpenetration with the rough surfaces ($h_{\mathrm{Tool}}$ and $h_{\mathrm{Def}}$).
Torques also arise due to tangential contact forces, with $\boldsymbol{\tau}^{(\mathrm{DEM})}$ accounting for particle–particle interactions and $\boldsymbol{\tau}^{(\mathrm{surface})}$ representing particle–surface interactions.
Damping terms are applied to the translational and rotational motion, denoted by $\eta$ in Equations \ref{eq:dem-linear} and \ref{eq:dem-rotation}, in order to include energy dissipation. Although lubricant fluid dynamics is not explicitly included, the purpose of such damping terms is to mimick viscosity effects acting on particles. \\

\begin{figure}[H]
	\centering
	\includegraphics[width=0.5\textwidth]{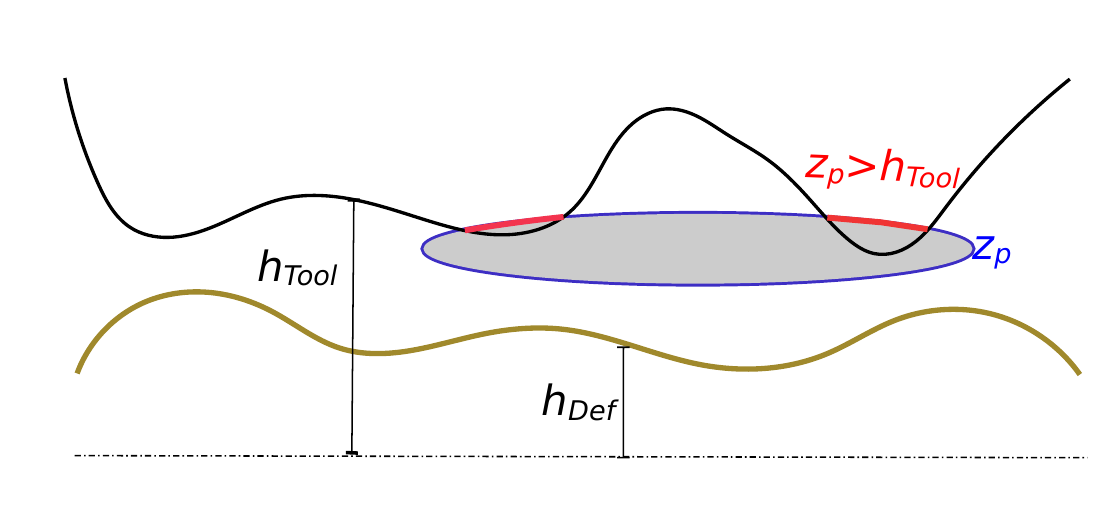}
	\caption{Schematic illustration (2D) showing the classification of a particle as \emph{blocked}. Spatially distinct clusters of surface points are identified where the particle surface height $ z_p(x, y) $ exceeds the local tool surface height $ h_{\text{Tool}}(x, y) $. A particle is labeled as blocked if more than three such clusters are in contact with the tool (in 2D it would be blocked with only two clusters as depicted in the figure).
	}
	\label{fig:surface-modification}
\end{figure}

\textbf{Step 2 -- Update the gap profile:}
The contact mechanics problem is solved using BEM. This assumes a low enough sliding velocity to have quasi-static conditions and justifying for a static resolution.
This classically demands a rough-rigid to be pressed against a flat-deformable surface \cite{Bonnet1999, PolonskyKeer1999}, using the plastic-saturation method that limits the contact traction below a critical value, $p_{crit}$ \cite{ref:pressure_saturation, ref:Weber2018}. This procedure yields the traction and separates the total displacement $u$ into its elastic and plastic components, denoted by $u_e$ and $u_p$, respectively. As was done in \cite{pundir}, our approach substitutes the rough-rigid profile with the gap $g$ separating the surfaces. Without particles, the gap is defined as
\begin{equation}
g = h_{\mathrm{Tool}} - h^0_{\mathrm{Def}}.
\end{equation}
Here, $h^0_{\mathrm{Def}}$ is the initial profile of the lower deformable surface.
However, particles blocked between the two surfaces will fill the gaps, allowing them to sustain normal loads. Furthermore, free particles should not transfer stress through the interface. Therefore, rigid particles are classified as \emph{free} or \emph{blocked}. A particle is assumed \emph{blocked} if at least three distinct areas of contact with the tool can be identified. A 2D schematic illustration is shown in Figure \ref{fig:surface-modification}.
If a particle is identified as \emph{blocked} within the interface, it can support a portion of the applied normal load. To let the BEM contact solver capture the mechanical influence of such particles, we modify the tool profile at the positions where the blocked particles are found to \textit{make them effectively part of the top surface}.
If $ \mathcal{P} $ denotes the set of all blocked particles, then for each particle $ j \in \mathcal{P} $, with its surface profile being $z_p$ the modified tool surface height $ h_{\text{Tool}}^\ast$ is:
\begin{equation}
	h_{\text{Tool}}^\ast(x, y) = \min \left( h_{\text{Tool}}(x, y), \min_{j \in \mathcal{P}} z_p(x, y) \right)
	\label{eq:tool_modified}
\end{equation}
It allows to define the gap profile including the presence of blocked particles:
\begin{equation}
	g^\star = h_{Tool}^\ast - h^0_{\mathrm{Def}}
\end{equation}
This gap will be used as the rough-rigid profile to be pressed against a flat deformable surface in the classical Fourier-BEM solver, as described in what follows.\\

\textbf{Step 3 -- Normal contact problem:} The BEM formulation relies on energy minimization principles applied to elasto-plastic half-space interfaces. This leads to the following inequality-constrained minimization problem:
\begin{equation}
\min_{
u \in \mathcal{K}} \frac{1}{2}\int_{S} u_e \cdot t_c(u_e), dS
\label{eq:lagrangian}
\end{equation}
where $\mathcal{K}$ is the set of admissible displacements satisfying the following constraints:

\begin{equation}
u \in \mathcal{K} \equiv 
\left\{
\begin{array}{ll}
g^\star - u_e - u_p \geq 0 &\text{Non-penetration condition} \\
0 \leq t_c(u_e) \leq p_{crit} &  \text{Admissible contact traction} \\
(g^{\star} - u_e - u_p)\,t_c(u_e) = 0 & \text{KKT complementarity condition \cite{Wright2006}} \\
p = \frac{1}{A_0} \int_S t_c(u_e)\,dS & \text{Equilibrating imposed pressure}
\end{array}
\right.
\end{equation}

where $p$ is the imposed overall scalar pressure, and $t_c(u_e)$ is the contact traction resulting from the elastic displacement $u_e$. The functional $t_c$ incorporates the effective Young’s modulus $E^\star$, which is set to 55 GPa to represent an aluminum sheet interfacing with a cast iron tool. The saturation pressure $p_{\text{crit}}$ is set to 340 MPa.
This problem is solved in practice by following the classical approach \cite{ref:pressure_saturation} implemented in the in-house open-source software \textit{Tamaas}\cite{Frerot2020}, which will deliver the elastic displacements and tractions as well as the cumulated plastic displacement, $u_p$. 

Finally, this solution has to be mapped back to the original rough-rigid against rough-deformable profiles, which is done by following the procedure described in \cite{pundir}.
Because the top surface is rigid, all the displacement $u_e$ and $u_p$ are attributed to the lower-deformable surface.  

From now on, one iteration is fully defined and comprises surface sliding, particle dynamics (DEM), gap update, and contact mechanics resolution (BEM). As stated earlier, it is repeated sequentially until the predefined sliding distance $\Delta x = 2 L$ is reached. Using this method, we can study the contact area covered by lubricating particles ($A_p$) and its relationship with system parameters such as normal pressure, sliding velocity, and particle density $\beta$. This analysis leads to a governing equation, described in detail in Section \ref{sec:results}, that provides a local friction coefficient $\mu$ to the macroscale model.

%




\section{Mesoscale model results}\label{sec:results}
As mentioned in Section \ref{sec:stip-draw}, experiments have highlighted the presence of debris flakes during strip-draw friction tests. To capture their effect on macroscopic friction, a simple multiscale friction model was proposed where the mesoscale frictional behavior of third-body particles blocked between rough surfaces is integrated into the macroscopic model described in Section \ref{sec:macro-num-model}. The coupling issued in Section \ref{sec:coupling-bem-dem} needs $A_p$, the area lubricated by wear particles, in Equations \ref{eq:mu} and \ref{eq:mu2} to compute an effective friction coefficient. 
In this section, we discuss the evaluation of $A_p$ from ensemble averages of mesoscale simulations, which will drive us to an analytical expression of the parameters controlling $A_p$.

\subsection{Surface profiles}
The averaged power spectral density (PSD) of the aluminum surface measured before the test, along with that of the tool (see Section~\ref{sec:setup-meso-confocal}), was used to generate a self-affine periodic surface, following the method outlined in the appendix of \cite{trapped}.
To account for the influence of different initial surface conditions, several additional deformable (aluminum) rough profiles were generated, with a shortest-to-longest cutoff wavelength ratio ranging from 1/100 to 4/100 and a Hurst exponent of 0.8.
The Hurst exponent was selected such that the resulting surfaces reproduced the averaged PSD slope of the measured aluminum surfaces presented in Section~\ref{sec:setup-meso-confocal} (see Figure~\ref{fig:exp_PSD}).
The results presented here are averaged over these initial configurations.
%
%

\begin{figure}[H]
	\centering
	a\includegraphics[width=0.4\textwidth]{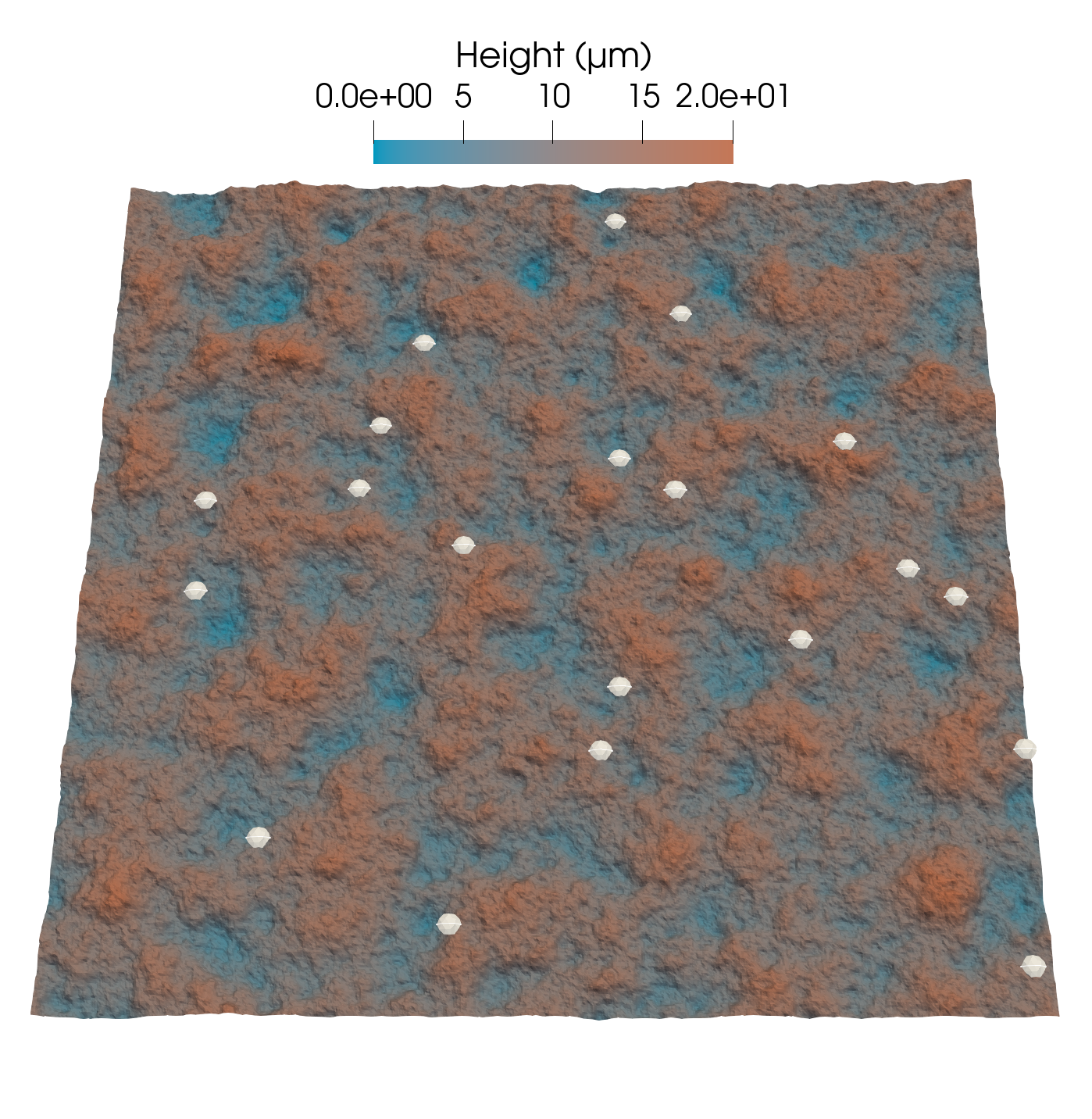}
    	b\includegraphics[width=0.4\textwidth]{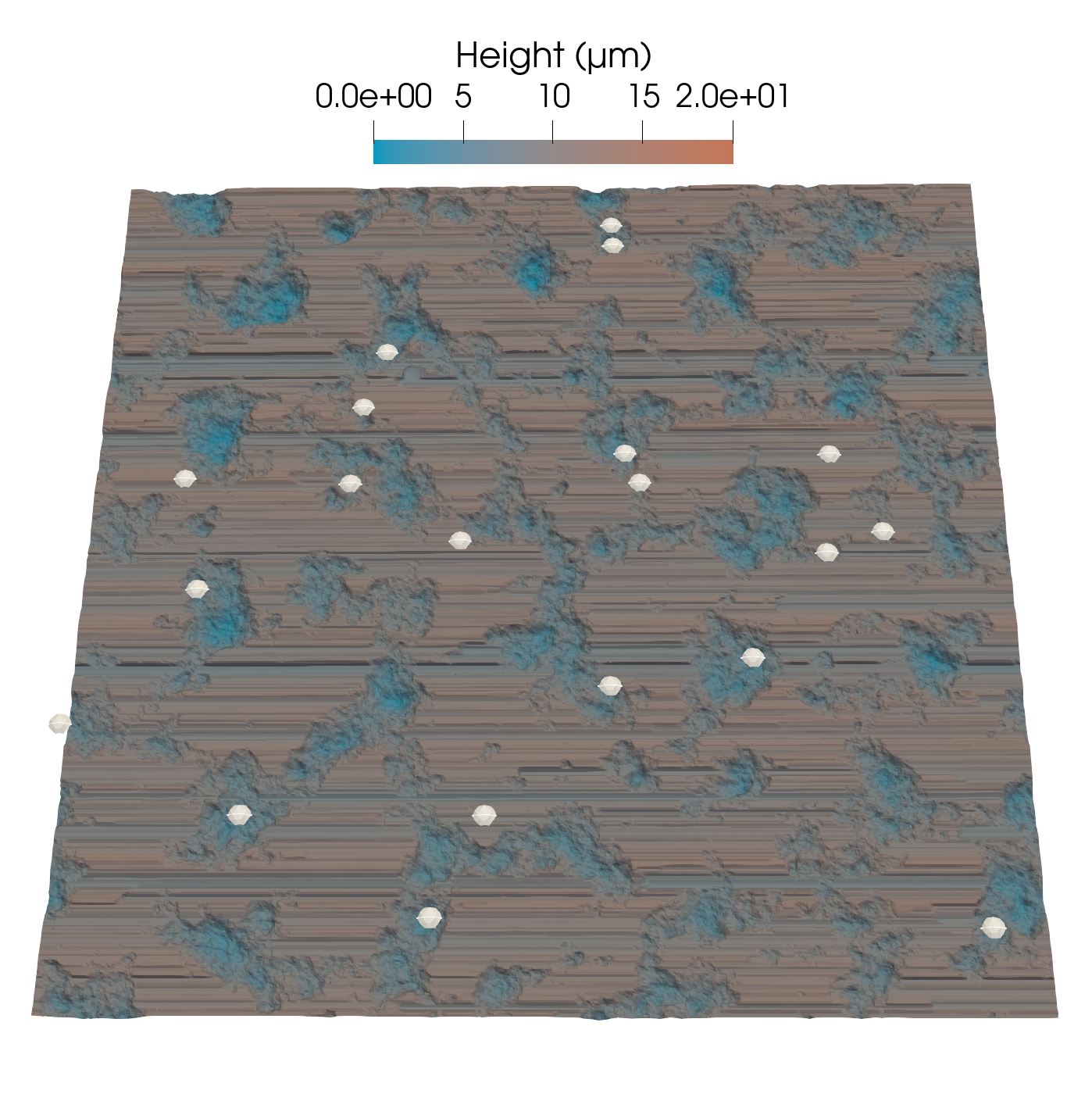}
	\caption{(a) Initial distribution of 20 oblate spheroidal particles, each with a diameter of 12\,\textmu m, on a deformable aluminum surface. (b) Distribution of particles on the deformed surface after 90\,mm of horizontal sliding under 10\,MPa normal pressure. Domain size: $500 \times 500\, \text{  \textmu m}^2$. The upper surface (tool) is omitted for clarity.}
	\label{fig:initial_part}
\end{figure}

The SEM images of the debris in Figure \ref{fig:SEM_image} show flake-shaped particles. For simplicity, the particles were modeled as thin rigid oblate spheres trapped between rough surfaces with no friction from particle--particle or particle--surface contacts. An example of the particle placement of 20 particles is shown in Figure \ref{fig:initial_part}, where the upper surface (tool) is not displayed in the figure, for clarity reasons. Figure~\ref{fig:initial_part}(b) also depicts the deformable surface after 90 mm of sliding under 10 MPa. Although our initially generated periodic self‑affine surfaces were not identical to the experimentally measured surfaces, the simulated smoothing and the horizontal grooves produced by the rigid tool in the sliding direction align with the experimental observations shown in Figure \ref{fig:confocal_PSD}.

\subsection{Measure of the evolution of $A_p$}\label{sec:measure-ap}

As noted in Section \ref{sec:coupling-bem-dem}, we solve the normal contact problem while explicitly accounting for blocked particles. The BEM solver provides the traction field, indicating where contact occurs. In addition, the positions of blocked particles are known at all times during the simulation. Therefore, the sub-areas where traction is positive (i.e., in contact) and a blocked particle is located underneath will contribute to $A_p$, as they represent the portion of the contact area supported by blocked particles.

\begin{figure}[H]
	\centering
	\includegraphics[width=0.55\textwidth]{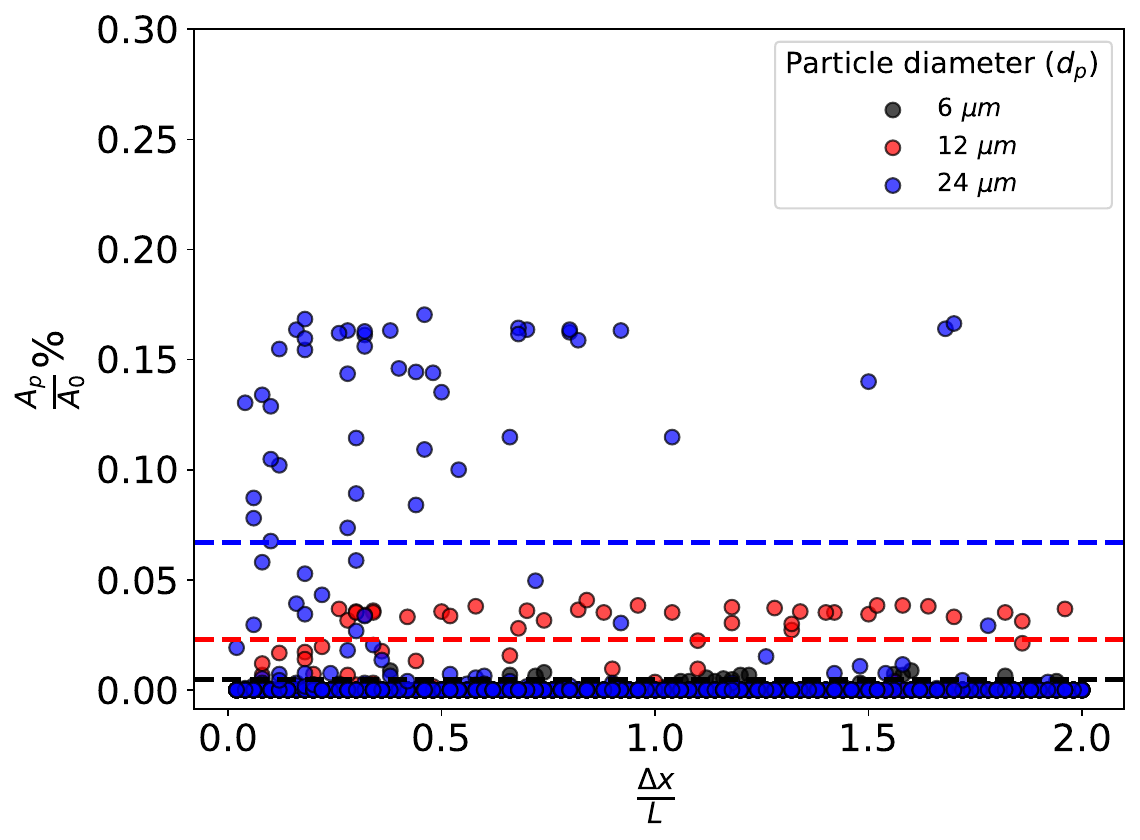}
	\caption{The distribution of $A_p/A_0$ (expressed in percent) as a function of the sliding distance $\Delta x/L$, is shown for a normal pressure of $10$ MPa and for three different particle diameters. Each simulation involves 20 particles placed on a periodic surface with dimensions $500 \times 500\, \text{  \textmu m}^2$.
Many of the points overlap when $A_p/A_0 = 0$. The dashed horizontal lines indicate the ensemble average $\bar{A_p}$ as described in Section \ref{sec:measure-ap}. The enforced sliding velocity is $50$ m/s.}
	\label{fig:scatter}
\end{figure}

Figure \ref{fig:scatter} shows $A_p$ as a function of the sliding distance for different particle diameters. The simulations include 20 particles placed between two rough and periodic surfaces of dimensions $500 \times 500\, \text{  \textmu m}^2$.
At first glance, the evolution of $A_p$ in Figure \ref{fig:scatter} appears random. This arises because the case where a particle is stuck in between the surfaces is a rather rare event, thus having most of the time no particle in contact.
However, it is expected that $ A_p $ increases together with particle radius, since larger particles have to cover a larger contact area. In a more subtile way, dynamical effects shall appear because particles falling into surface valleys, to become deactivated, have to be impacted by the sliding surfaces. 


\begin{figure}[H]
	\centering
	\includegraphics[width=0.5\textwidth]{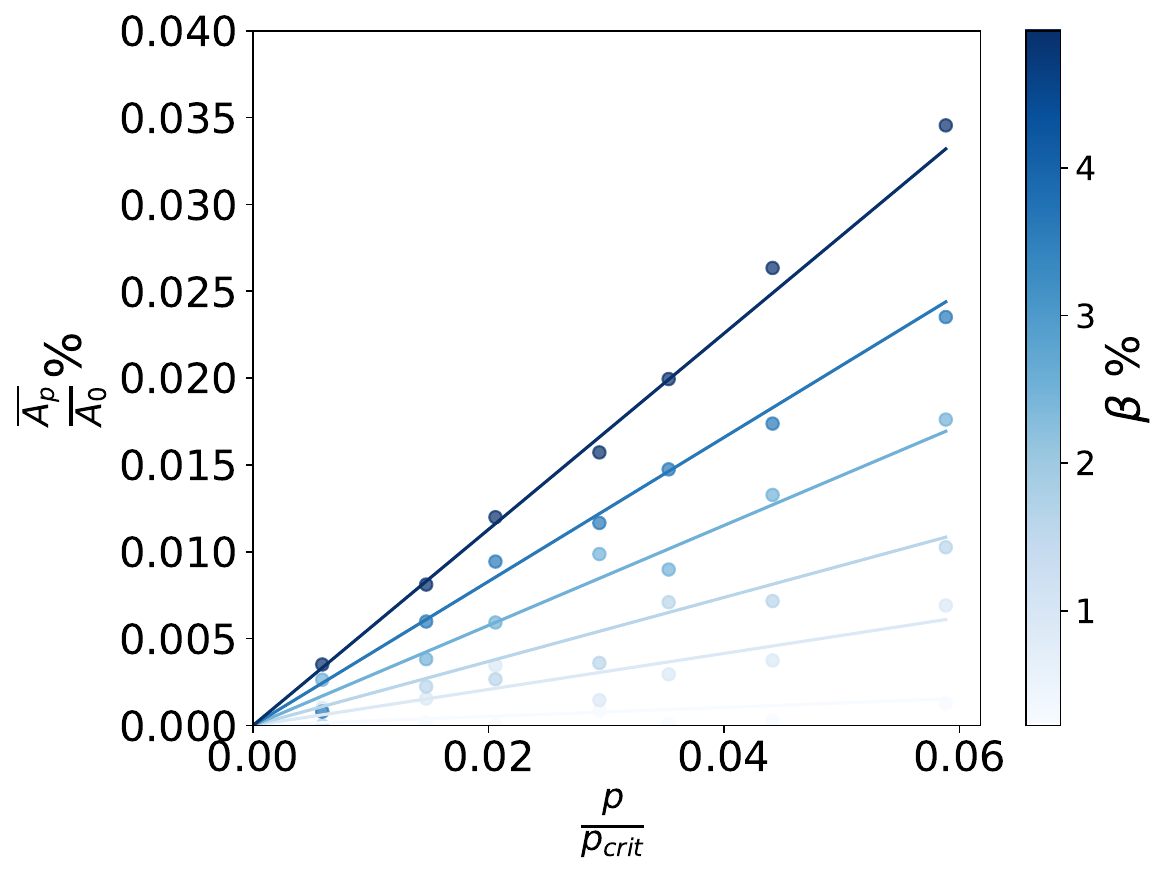}
	\caption{The contact area involving trapped particles, $\bar{A_p}$, normalized by the apparent contact area $A_0$, is shown (expressed in percent) as a function of applied normal pressure for different particle area densities, at a sliding velocity of 50\,m/s. The fitted lines are proportional to $\beta p$, demonstrating a linear relationship between $\bar{A_p}$ and both the particle area density $\beta$ and the applied normal pressure $p$.
    }
	\label{fig:Ap}
\end{figure}

At the macroscopic scale, because the time-scale is much longer, no fluctuations should be observed, and an event-based description has to be modified. The time scale shift can be accounted for with an averaged measure of $A_p$, made over the total sliding distance:

\begin{equation}
\left<A_p\right> = \frac{1}{\Delta_x}\int_0^{\Delta_x} A_p(x) dx \simeq \frac{1}{\Delta_x} \sum_i^n A_{p,i} \delta x = \frac{1}{n} \sum_i^n A_{p,i},
\end{equation}
where $\delta x$ is the sliding increment and $\Delta x$ is the total sliding distance, $n$ is the number of DEM steps and $A_{p,i}$ is the 
measure of $A_p$ at time-step $i$. 
An ensemble-averaged $\bar{A}_p$ was then obtained by averaging $\left<A_p\right>$ over three simulations with different initial random distributions of particles. Figure \ref{fig:Ap} presents the normalized particle contact area, $\bar{A}_p$, as a function of particle density and applied normal pressure. The normal pressures investigated are of the same order as those used in the experimental strip-draw friction tests (see Figure \ref{fig:combined_experiments}). 
The results in Figure \ref{fig:Ap} show that $\bar{A}_p$ increases linearly with the density of the particles and the normal pressure, allowing us to propose the following relation: 
\begin{equation}
	\frac{\bar{A_p}}{A_0} = \alpha \cdot \beta \frac{p}{E^\star},
	\label{eq:final-A-p}
\end{equation}
where $\alpha$ is a dimensionless parameter that is neither dependent on the applied pressure nor on $\beta$ the concentration of particles. As will be seen later, it depends on the sliding velocity and on geometrical factors such as the average particle-surface orientations. Our interpretation is that this parameter reflects the inertial effects that prevent particles to deactivate into valleys.

Having obtained the formulation for $\bar{A}_p$, the saturation threshold $\beta^{\mathrm{thres}}$ can be derived from Equations \ref{eq:final-A-p} and \ref{eq:mu2}. By imposing $A_c \ge A_p$, we then find:
\begin{equation}
  \beta^{\mathrm{thres}} = \frac{\kappa}{\alpha\,h'_{\mathrm{RMS}}}\,.
  \label{eq:b_thres}
\end{equation}

\subsection{Particle sizes and available gap}

\begin{figure}[H]
	\centering
	\includegraphics[width=0.7\textwidth]{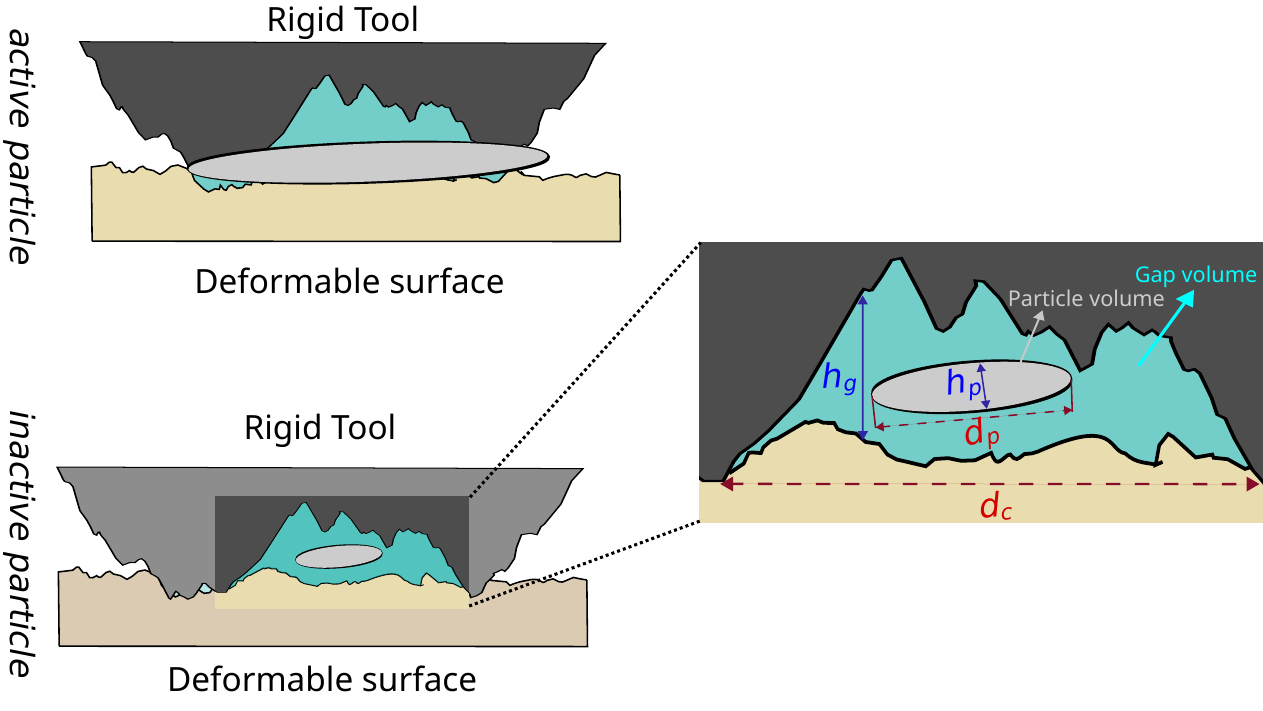}
	\caption{Schematic illustration of a rigid particle positioned between two rough surfaces. The particle may settle into surface valleys depending on its size and the local surface topography.  The schematic illustrates the hypothesis that the average contact area borne by the particles, $A_p$, is related by the ratio of the particle volume to the volume of the interfacial gap, as expressed in Equation \ref{eq:Ap_formula_hypothesis}.}

	\label{fig:Ap_formula_hypothesis}
\end{figure}

The dependency of $\bar{A}_p$ with $\beta$ and $p$ will be further
justified in this section. A necessary condition for a particle to migrate to surface valleys is that its dimensions fit in-between the contacting asperities. Figure \ref{fig:Ap_formula_hypothesis} illustrates a particle positioned between two rough surfaces. In the light of this argument, it is natural that one particle contributes to $\bar{A_p}$ with the ratio of the particle volume over the volume of the gap formed by the surfaces. Accordingly, we can write:
\begin{equation}
	\frac{\bar{A_p}(P)}{A_0} \propto
	\frac{N_p a_p}{A_0} \biggl(\frac{a_p \cdot h_p}{{\bar{d_c}}^2 \bar{h_g}}\biggr),
	\label{eq:Ap_formula_hypothesis}
\end{equation}
where $N_p$ is the number of particles, $\bar{d_c}$ is the average minimum distance between contact clusters, and $h_p$ and $\bar{h_g}$ are the particle height and the average gap height profile, respectively.
Both $\bar{d_c}$ and $\bar{h_g}$ are functions of the applied normal pressure, $p$. 
Within the range of applied loads, the mean gap height $\bar{h_g}$ is much larger (at least five times greater) than the thin particle height $h_p$.
In the studied cases, $h_p/\bar{h_g}$ was then considered a constant. This simplification would be invalid with thicker flakes or with a much smaller average gap (i.e., for smooth surfaces and/or critically large pressures). However, it means that for the present work, the vertical available space $h_p/\bar{h_g}$ is not an important factor.

Therefore, the dominant geometric factor governing whether a flake can settle into a surface valley is the ratio of its diameter to the distance between contact clusters, i.e., $\sqrt{a_p} / \bar{d_c}$.
Substituting Equation~\ref{eq:beta}, i.e., $\beta = N_p a_p / A_0$, into Equation~\ref{eq:Ap_formula_hypothesis} gives:
\begin{equation}
	\frac{\bar{A_p}}{A_0} \propto
	\beta \frac{a_p}{{\bar{d_c}}^2}
\end{equation}
%

\begin{figure}[H]
	\centering
	\includegraphics[width=0.9\textwidth]{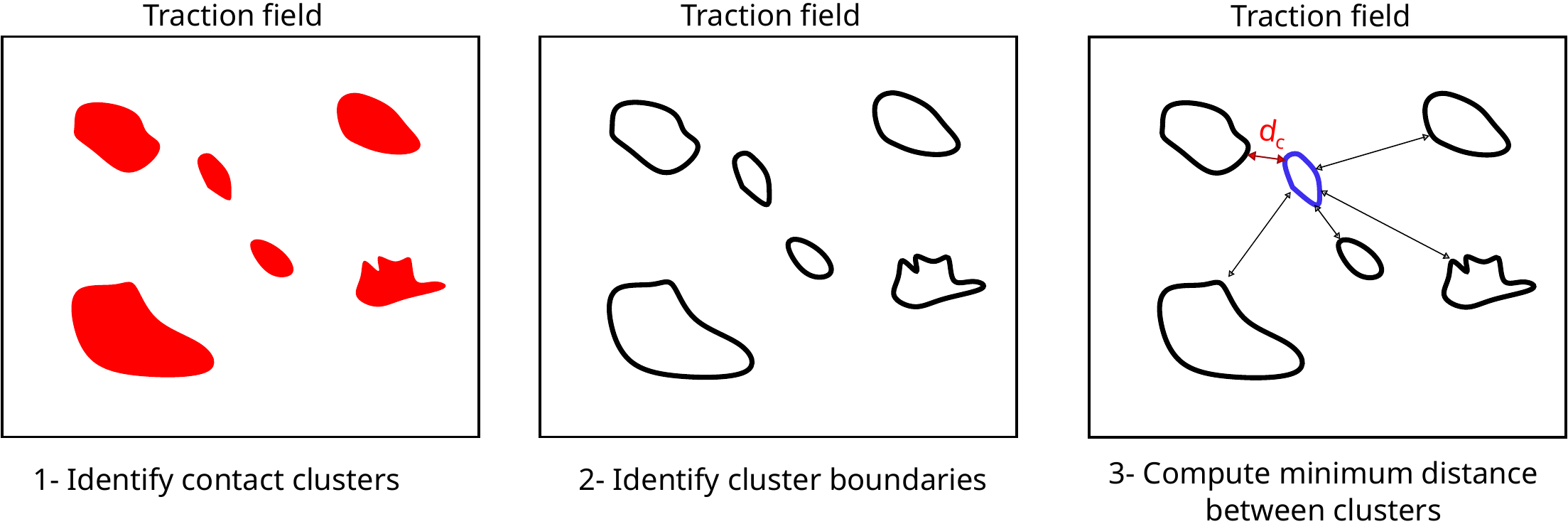}
	\caption{Schematic illustration of the procedure used to extract $d_c$ for a single contact cluster, defined as the minimum distance between the border of that cluster and the borders of its neighboring clusters.}
	\label{fig:Shematic_dc}
\end{figure}

\begin{figure}[H]
	\centering
	\includegraphics[width=0.55\textwidth]{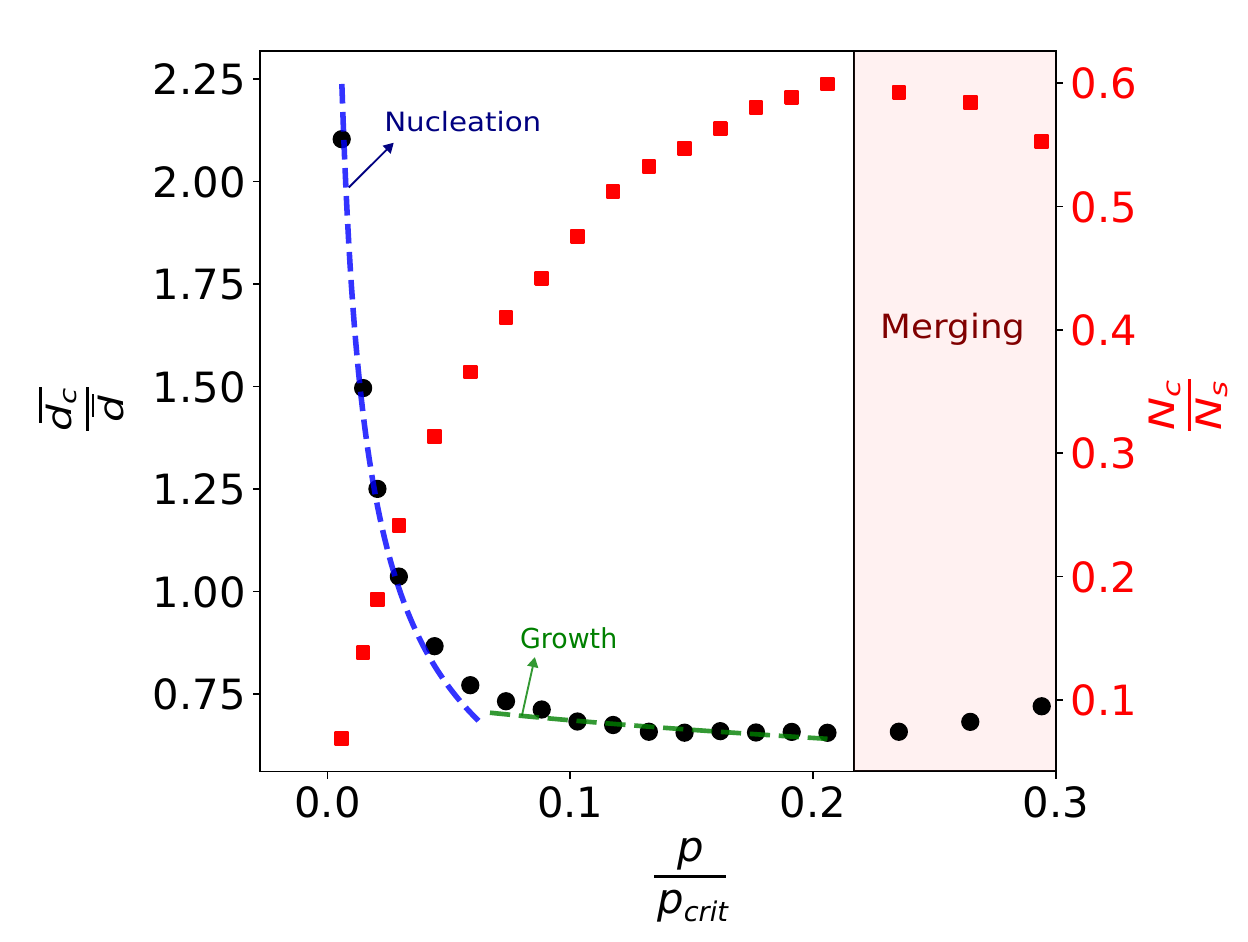}
	\caption{Left black y-axis: Evolution of $\bar{d_c}$, the average minimum distance between contact clusters, normalized by the mean center-to-center distance between summits prior to contact, defined as $\bar{d} = \frac{L}{\sqrt{N_s}}$, where $N_s$ is the number of summits in the initial gap between the rough surfaces, $h_{\mathrm{Tool}} - h_{\mathrm{Def}}$, and $L$ is the mesoscale domain length. The value of $N_s$ is obtained by identifying local maxima. The $\bar{d_c}$ values, extracted from simulation results, are fitted to the trends $\bar{d_c} \propto 1/\sqrt{p}$ and $\bar{d_c} \propto \bar{d} - C\sqrt{p}$, where $C$ is a fitting parameter defined in Appendices~\ref{sec:nucleation} and \ref{sec:growth}.
Right red y-axis: Evolution of the number of contact clusters, $N_c$, normalized by $N_s$.}
	\label{fig:dc}
\end{figure}

As illustrated in Figure \ref{fig:Shematic_dc}, $\bar{d_c}$ can be computed from the previously presented mesoscale simulations, with the following procedure: after identifying the contact clusters and their boundaries with a classical coloring algorithm \cite{Azriel1966}, every cluster will compute the minimum distance with its neighbors (distance separating boundaries). The minimum distance is then averaged over all clusters to obtain a measure of $\bar{d_c}$. 
Figure \ref{fig:dc} shows for both the number of contact clusters and the average distance between contacts areas ($\bar{d_c}$), their evolution when normal load increases. 
%
%
Initially, when rough surfaces are first brought into contact under low to moderate loads, micro-contacts (asperity tips) nucleate at isolated points. It can be observed that $\bar{d_c} \propto \frac{1}{\sqrt{p}}$, which characterizes the \textit{nucleation} phase. At higher pressures, the contact clusters will grow substantially, which eventually affects the spacing between neighboring asperities in contact. Then, the behavior is better captured by the relation $\bar{d_c} \propto \bar{d} - C\sqrt{p}$, corresponding to the \textit{growth} phase. Detailed derivations of the asymptotes in these regimes are provided in Appendices \ref{sec:nucleation} and \ref{sec:growth}, respectively.
It is important to note that our simulations predominantly lie within the nucleation regime of contact clusters ($\frac{p}{p_{\mathrm{crit}}} < 0.09$), which justifies the use of the relation $\bar{d_c} \propto \frac{1}{\sqrt{p}}$ in the formulation of $\bar{A_p}$, yielding:

\begin{equation}
	\bar{A_p} \propto
	\beta p,
\end{equation}
which is the same as equation \ref{eq:final-A-p}, and demonstrates that the key factor in the \textit{nucleation} phase is the ratio between $\bar{d_c}$ and particles' characteristic length $\sqrt{a_p}$.

%
%

	%
%

%

%

\subsection{Inertial effects and $\alpha$}

\begin{figure}[H]
	\centering
    \includegraphics[width=0.35\textwidth]{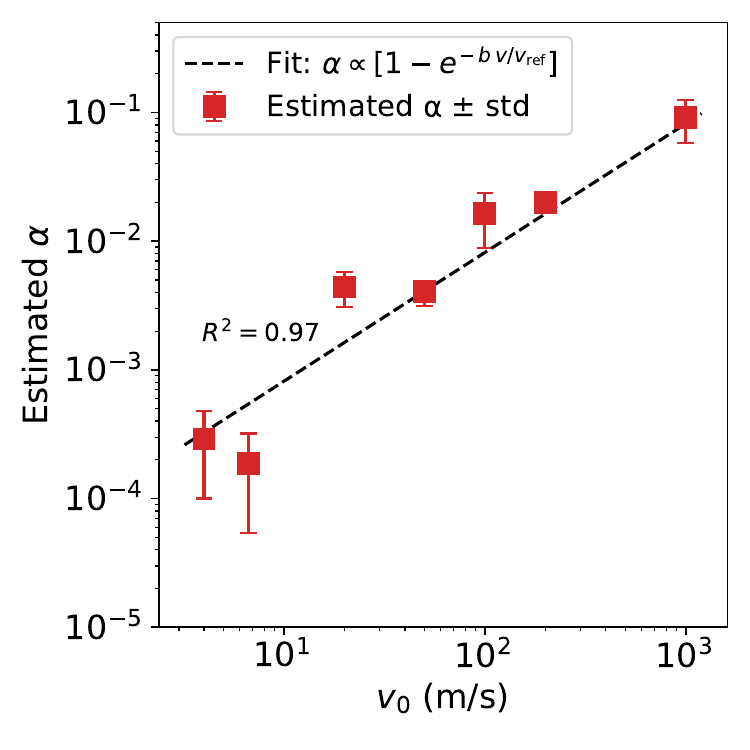}
	\caption{The estimated dimensionless coefficient $\alpha$ (the fitted parameter in Equation \ref{eq:final-A-p} for constant velocity) are plotted as functions of the sliding velocity $v_0$. Each value of $\alpha$ was determined over a range of normal pressures from 2 to 30 MPa. Error bars indicate the standard deviation of $\alpha$ across three different initial random particle distribution. The dashed line shows the best‐fit, with $b$ and $v_{ref}$ as the fitting parameters. }

	\label{fig:alpha}
\end{figure}

As was mentioned previously, $\alpha$ in Equation (\ref{eq:final-A-p}) is expected to depend on particle-surface orientations and on the sliding velocity. Note that $\bar{A_p}$ is computed from simulations with a fixed sliding distance and with the same rough surfaces. In this context, the sliding velocity constrains the time available for particles to find their way into valleys and deactivate. Consequently, the probability of having active particles, i.e., particles bearing the applied normal load, should increase with velocity. This is confirmed by the results shown in Figure \ref{fig:alpha}, which illustrates the dependence of $\alpha$ on the sliding velocity $v_0$ (the standard deviation is obtained from the three different initial random distribution of particles). It can be seen that $\alpha$ increases with sliding velocity following an exponential saturation law. We used the estimated values of $\alpha$ directly in our macroscale model.

\begin{figure}[H]
	\centering
	\includegraphics[width=0.45\textwidth]
    {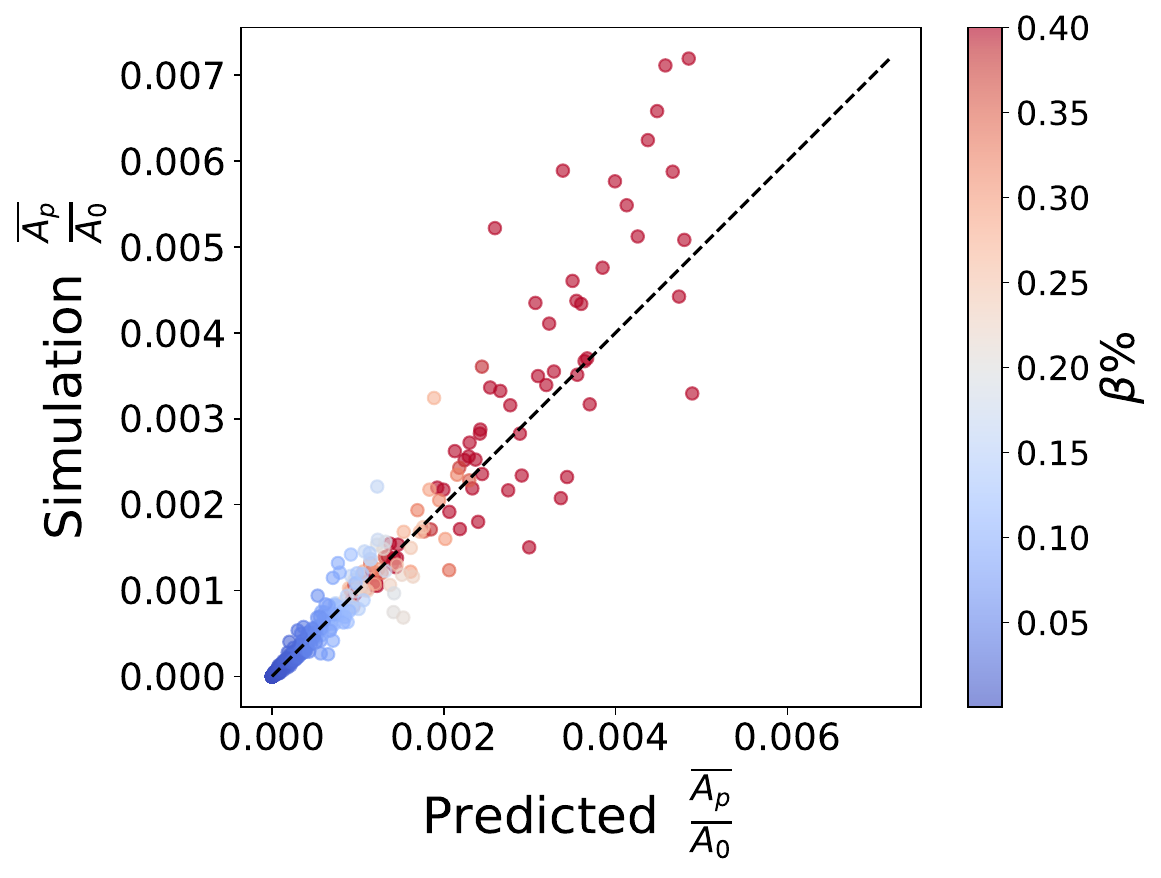}
	\caption{Comparison between the proposed function for the average trapped-particle contact area, $\bar{A_p}$ (defined in Equation \ref{eq:final-A-p}), and the simulation results. The data points are color-coded by $\beta$ value.}
	\label{fig:Ap_comp}
\end{figure}

A comparison between $\bar{A_p}(p, \alpha(v_0), \beta)$ predicted by equation \ref{eq:final-A-p} and the results of the mesoscale DEM simulation in Figure \ref{fig:Ap_comp} shows that this model is accurate for a wide range of particle density, pressure and velocities.
Note that Equation \ref{eq:final-A-p} should be used with caution at higher particle area densities, $\beta$, as the linear relationship between $A_p$, $\beta$, and the applied normal pressure breaks down when $\beta$ becomes large.
A high $\beta$ indicates that particles occupy a larger fraction of the nominal surface area. This can occur in two ways. First, when particle diameters are large, the particles tend to remain on top of the asperities, masking the underlying surface roughness. As a result, the contact behavior deviates from that of self-affine surfaces, where the contact area increases linearly with the normal pressure. Second, a high $\beta$ can result from a greater number of particles. In this case, particle–particle interactions become significant and can influence their mechanical evolution. This again deviates from the assumptions behind the linear formulation in Equation \ref{eq:final-A-p}. In this work, we assume a low particle density, where the linear formulation remains sufficient to account for particle effects at the macroscale, as will be discussed later.
%

\section{Multiscale results}

In this section we present the multiscale results, where the mesoscale expression for $\bar{A}_p$ is integrated into the the Equation \ref{eq:mu} defining the local friction coefficient. 
As already presented in Section \ref{sec:macro-num-model}, a rigid tool is pressed 
with a constant pressure onto a one-dimensional elastic deformable bar (an aluminum sheet in our case) which is pulled at a fixed sliding velocity of $0.08$ m/s from its extremity (see Figure \ref{fig:exp}(b)). This setup serves as a simplified representation of the strip-draw friction tests conducted experimentally, allowing to compare predictions of global traction resistance $F_S$ with experimental ones.
%
%
%
The averaged global friction coefficient, denoted as $\bar{\mu}$, is computed as:
\begin{equation}
	\bar{\mu} = \frac{F_S}{F_N},
\end{equation}
where $F_S$ is the total friction force and $F_N$ is the total normal force acting on the bar.  
To ensure consistent comparison with experimental results, $\bar{\mu}$ is extracted from the steady-state region near the end of the friction coefficient–sliding distance curve, with the total sliding distance fixed at 140\,mm for all simulations.
Remarkably, the time evolution of the particle area density, $\beta(x,t)$, during the strip-draw test can also be obtained. It depends on $\zeta$, the particle migration–diffusion coefficient (accounting for transport and adhesion to the tool), and on the particle generation rate $K$, also known as the Archard coefficient.
%

\begin{figure}[H]
	\centering
	a\includegraphics[width=0.8\textwidth]{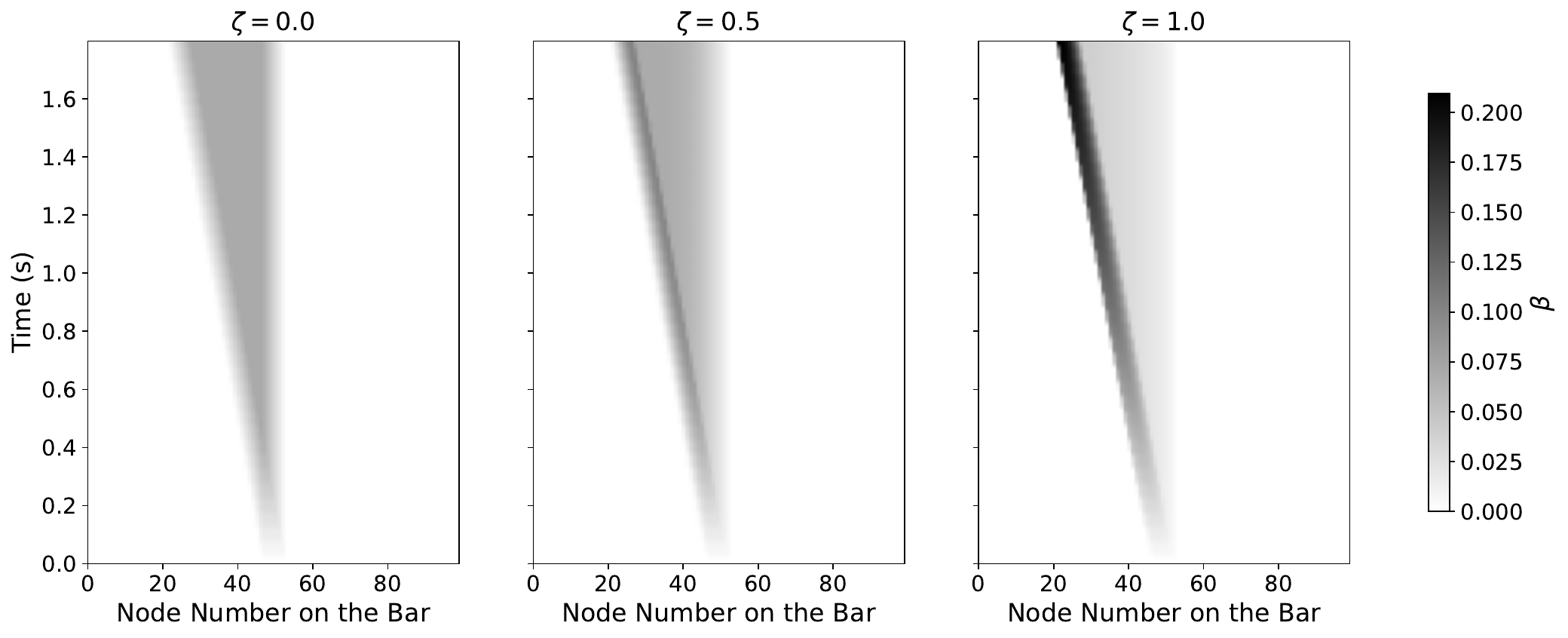}

	b\includegraphics[width=0.5\textwidth]{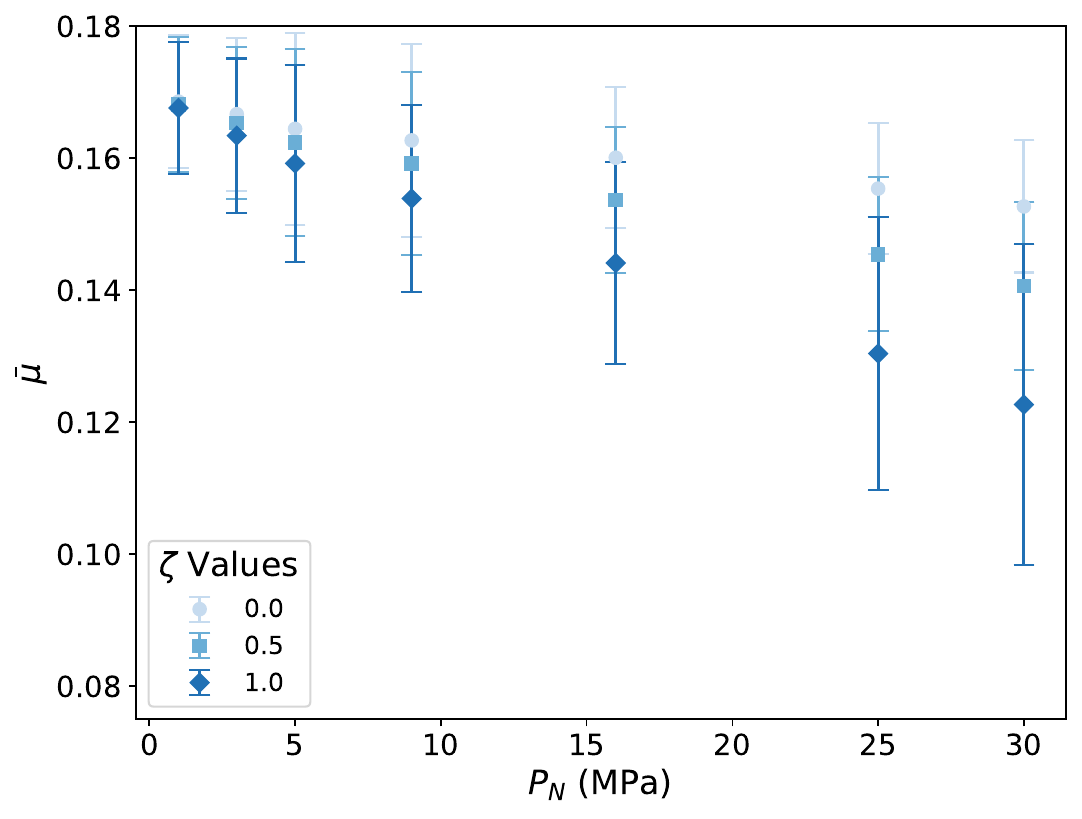}
	\caption{
		(a) Time and position-dependent evolution of the particle area fraction $\beta$ on the aluminum sheet and the macroscopic friction coefficient $\Bar{\mu}$, shown for varying particle flux values $\zeta$ at $P_N=30$ MPa. Increase in $\beta$ reflects progressive particle generation due to tool–sheet interaction. (b) Macroscopic friction coefficient vs normal applied load for different particle flux values $\zeta$. Higher $\zeta$ leads to greater particle accumulation beneath the tool, resulting in a lower macroscopic friction coefficient. The parameters used in the simulation are: Tool length = 35 mm, and $K = 1.0 \times 10^{-4}$.
	}

	\label{fig:macro_zeta}
\end{figure}

Figure~\ref{fig:macro_zeta}(a) shows the time–space evolution of the particle distribution, $\beta$, for different values of the diffusion, $\zeta$. $\zeta = 0$ corresponds to the case where all particles remain on the aluminum plate, while $\zeta = 1$ represents the case where all particles adhere to the tool.
In all cases, a "V"-shape can be observed. At a given time, the grayed horizontal region marks the sliding distance.
The right end of the "V"-shape is a vertical line marking the position of the first point on the bar that came into contact at the onset of the strip-draw test. On its left end, an inclined line is observed, whose slope is proportional to the sliding velocity and indicates the location of the tool.
Globally, the quantity of particles increases with time (or equivalently, with sliding distance). This trend is natural, as more wear particles are generated with increasing sliding distance.

The parameter $\zeta$ clearly influences the distribution of particles between the two lines.
As $\zeta$ increases, particles adhering to the tool eventually meet the newly generated particles, forming an accumulation zone. This leads to a higher particle area fraction, $\beta$, on the bar nodes beneath the tool and, consequently, to a lower macroscopic friction coefficient, as shown in Figure~\ref{fig:macro_zeta}(b).

\begin{figure}[H]
	\centering
	a \includegraphics[width=0.5\textwidth]{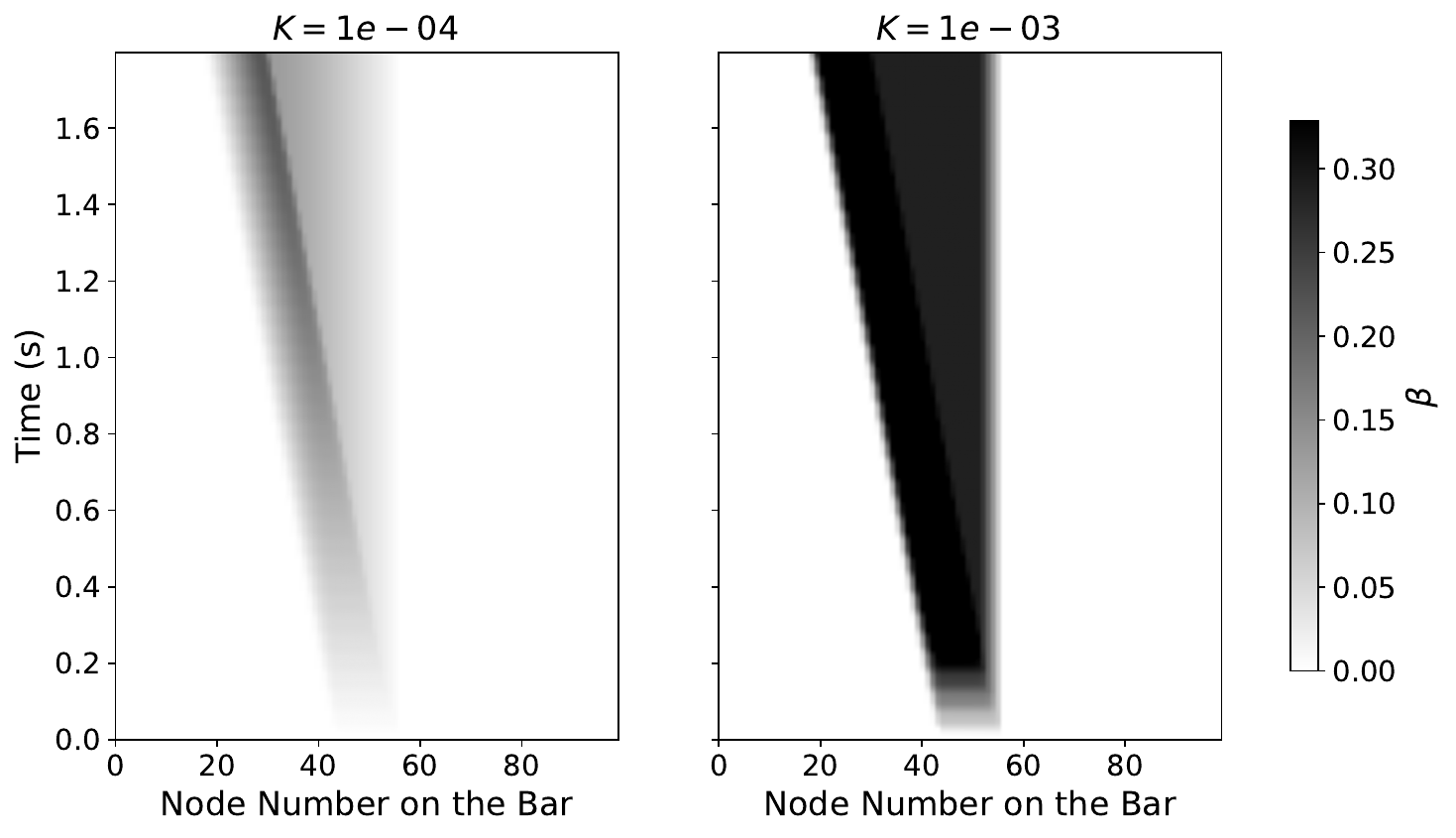}
	b\includegraphics[width=0.5\textwidth]{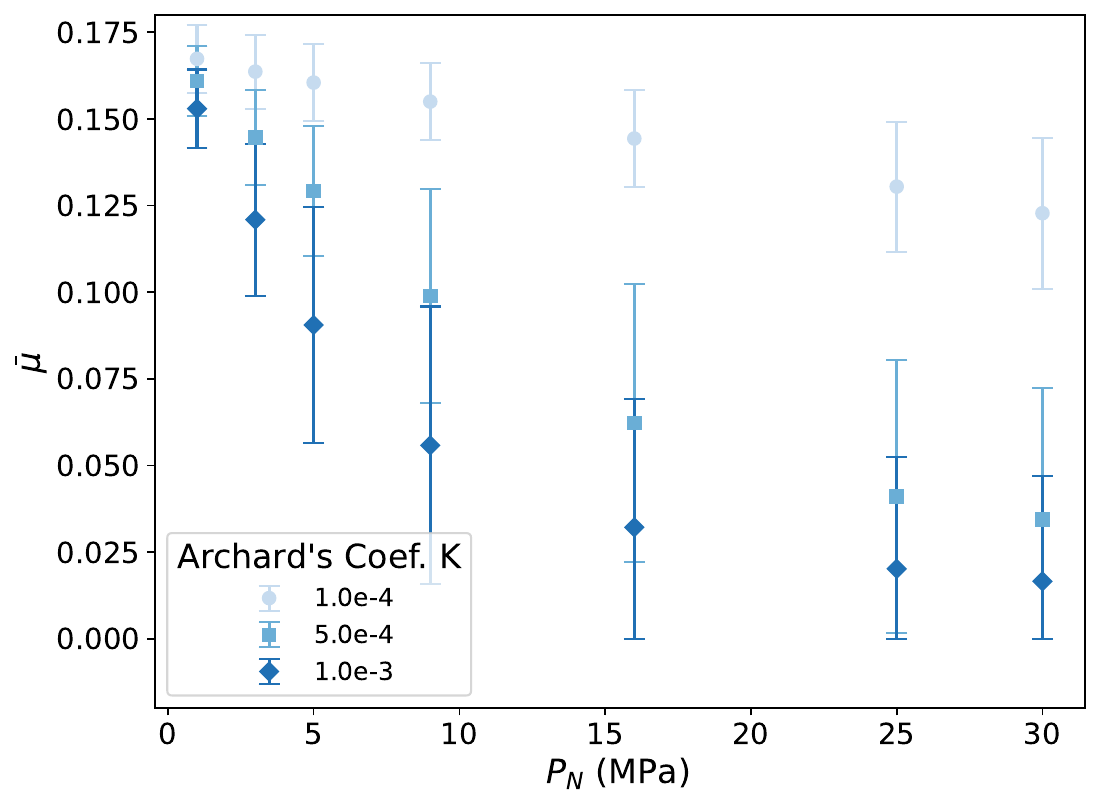}
	\caption{(a) Time and position-dependent evolution of the particle area fraction $\beta$ on the aluminum sheet and the macroscopic friction coefficient $\Bar{\mu}$, shown for varying Archard's wear coefficient values $K$ at $P_N=30$ MPa. Increase in $\beta$ reflects progressive particle generation due to tool–sheet interaction. (b) Macroscopic friction coefficient vs normal applied load for different Archard's wear coefficient values $K$. Higher $K$ leads to the generation of more wear particle, resulting in a lower macroscopic friction coefficient. Tool length = 70 mm, and $\zeta=0.0$}
	\label{fig:macro_archard}
\end{figure}

The other influencial parameter of the macroscopic model is Archard's wear coefficient, $K$.
Figure \ref{fig:macro_archard}(a) illustrates the spatial distribution of particles along the bar (aluminum sheet) over time for two different values of Archard's wear coefficient. Because the particle area fraction cannot exceed $\beta^{\mathrm{thres}}$, both the average $\beta$ and the macroscopic friction coefficient saturate once particle production becomes excessive. This effect appears at large values of $K$ in Figure \ref{fig:macro_archard}(a). Moreover, Figure \ref{fig:macro_archard}(b) shows that the saturation of $\beta$ leads to a nonlinear decrease in the friction coefficient with increasing applied normal pressure, $P_N$, a trend also observed in the experimental strip-draw tests presented in Section \ref{sec:stip-draw}. In this paper, we choose $\beta^{\mathrm{thres}}$ based on the simple fact that the particle contribution to the real contact area, $A_p$, cannot exceed the total contact area, $A_c$ (see Equation \ref{eq:b_thres}). Further research could investigate the physical mechanisms that limit particle generation and how they influence the macroscopic friction coefficient.


\begin{figure}[H]
	\centering
	a\includegraphics[width=0.7\textwidth]{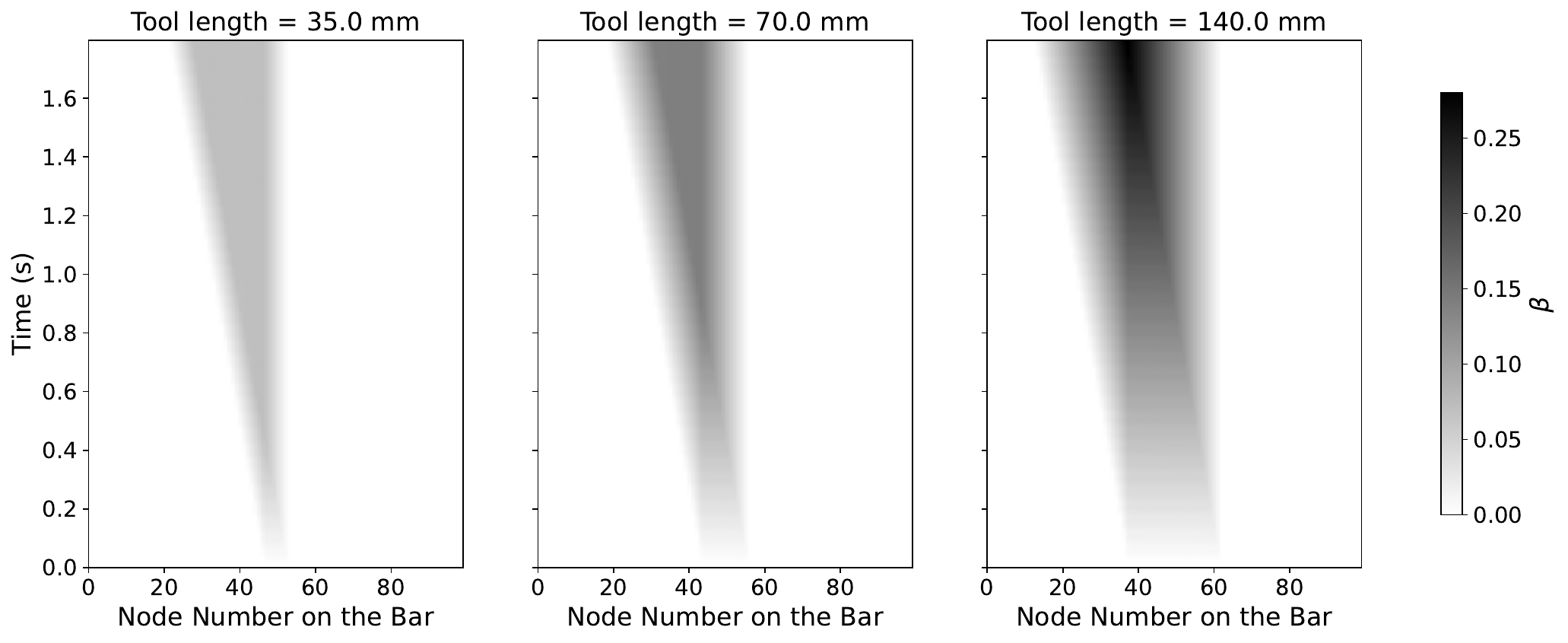} \\
	b\includegraphics[width=0.55\textwidth]{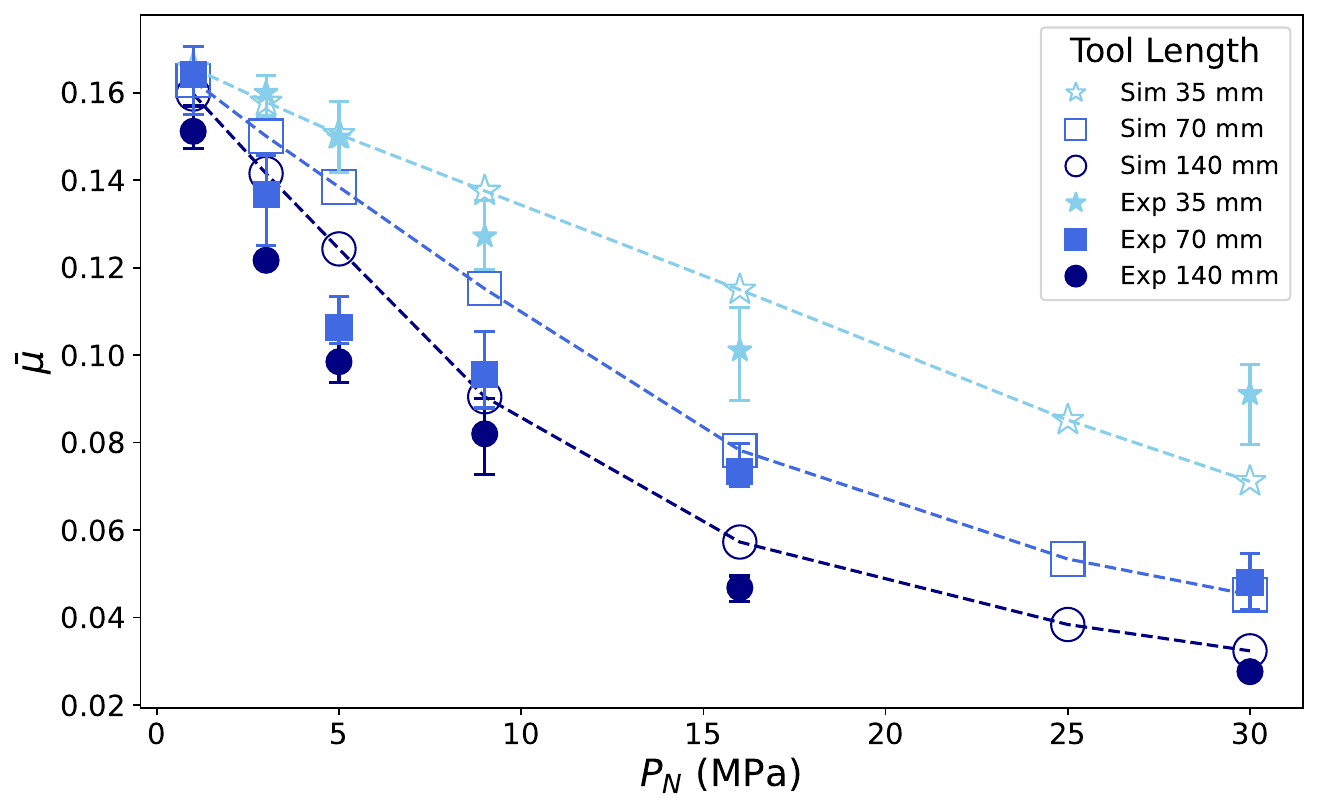}
	\caption{(a) Time and position-dependent evolution of the particle area fraction $\beta$ on the aluminum sheet and the macroscopic friction coefficient $\Bar{\mu}$, shown for varying tool length at $P_N=30$ MPa. An increase in $\beta$ reflects progressive particle generation due to tool–sheet interaction.  
		(b) Comparison between the experimentally measured friction coefficient from the strip-draw friction test of EDT Aluminum 5182, and the predicted values from the developed multiscale model. The figure shows the dependency of the macroscopic friction coefficient on tool size. Larger tool sizes result in lower friction coefficients due to increased particle accumulation. The width of the sheet was 35 mm in all tests. The parameters used in the simulation are: $\alpha$ was extrapolated from Figure~\ref{fig:alpha} using the pulling velocity $v_0 = 0.08~\text{m/s}$. The value $K = 1.0 \times 10^{-4}$ was chosen within the typical range of Archard’s wear coefficient for aluminum alloys, as shown in the Ashby diagram~\cite{ashby}. The only fitting parameter in the model was $\zeta = 0.2$.
	}
	\label{fig:exp_comparision}
\end{figure}

One of the key advantages of the developed multiscale model is its ability to capture the tool’s size-dependent behavior, as observed in Section \ref{sec:stip-draw}.
This is particularly significant for industrial applications, where friction tests can be conducted for various tool/plate sizes, depending the investigation goal or the machine dimensions. Our model predicts that the friction coefficient decreases with increasing tool size (see Figure \ref{fig:exp_comparision}), as more particles will accumulate under larger tools, therefore enhancing the lubrication effect. This behavior was confirmed experimentally, and our multiscale approach favorably compares with the experimental data.
This size-dependency prediction is crucial for scaling laboratory-scale friction tests to real-world industrial applications. Accounting for the tool size effect by means of lubricating wear particles, lets the model uncover the origins of the frictional behavior in practical scenarios, and enables better design and optimization of tribological systems.


\section{Conclusion}

This work presents a multiscale model for predicting friction behavior in metal forming, bearing systems, and other contact-based applications where third-body particles are present.  
We predict the macroscopic friction while accounting for the interface history at the mesoscale, incorporating particle dynamics, contact evolution, and wear-induced particle generation during sliding contacts.

At the mesoscale, a coupled BEM--DEM framework was used to simulate the interaction between particles and rough surfaces, with one of the surfaces deformable. This model enabled tracking particle motion between the surfaces and calculating key parameters, such as the contribution of blocked particles to the real contact area ($A_p$) and the evolution of contact clusters under varying normal pressures.

At the macroscale, simulations showed that increased particle accumulation, either from a higher fraction of particles adhering to the tool surface ($\zeta$) or from increased material wear (represented by Archard's wear coefficient, $K$), leads to a significant reduction in the macroscopic friction coefficient. Moreover, the model successfully captured the size dependency of the tool: larger tools accommodate more particles, resulting in enhanced lubrication and reduced friction. The predictions showed good agreement with experimental results, supporting the model’s robustness and practical relevance.

Future work may focus on implementing a more advanced friction model based on the rate-and-state friction law, as well as investigating the occurrence of stick--slip during sliding, the influence of particle shape, and particle deformation and fracture, to further improve the model’s accuracy and applicability.

\section*{Acknowledgements}
\paragraph{} The researchers express their gratitude to Innosuisse (Swiss Innovation Agency) for funding this project under grant number 100.777 IP-ENG and to Novelis for their support through a research agreement.

\newpage

\section*{Appendix}                
\addcontentsline{toc}{section}{Appendix}

\renewcommand\thesubsection{\Alph{subsection}}
\setcounter{subsection}{0}

\subsection{Influence of tool pad width on the friction coefficient}
\label{sec:pad-width}

To investigate the role of pad geometry in frictional performance during strip-draw testing, four tool pad configurations were selected. The pads were defined by their length~$\times$~width dimensions, with length corresponding to the sliding direction. The tested sizes were 140~$\times$~70 mm, 140~$\times$~35 mm, 70~$\times$~35 mm, and 35~$\times$~35 mm.
This setup enabled a systematic comparison of the influence of pad width versus pad length, allowing assessment of whether the total contact area or one specific dimension is the dominant factor in frictional behavior.

\begin{figure}[H]
	\centering

	a\includegraphics[width=0.45\textwidth]{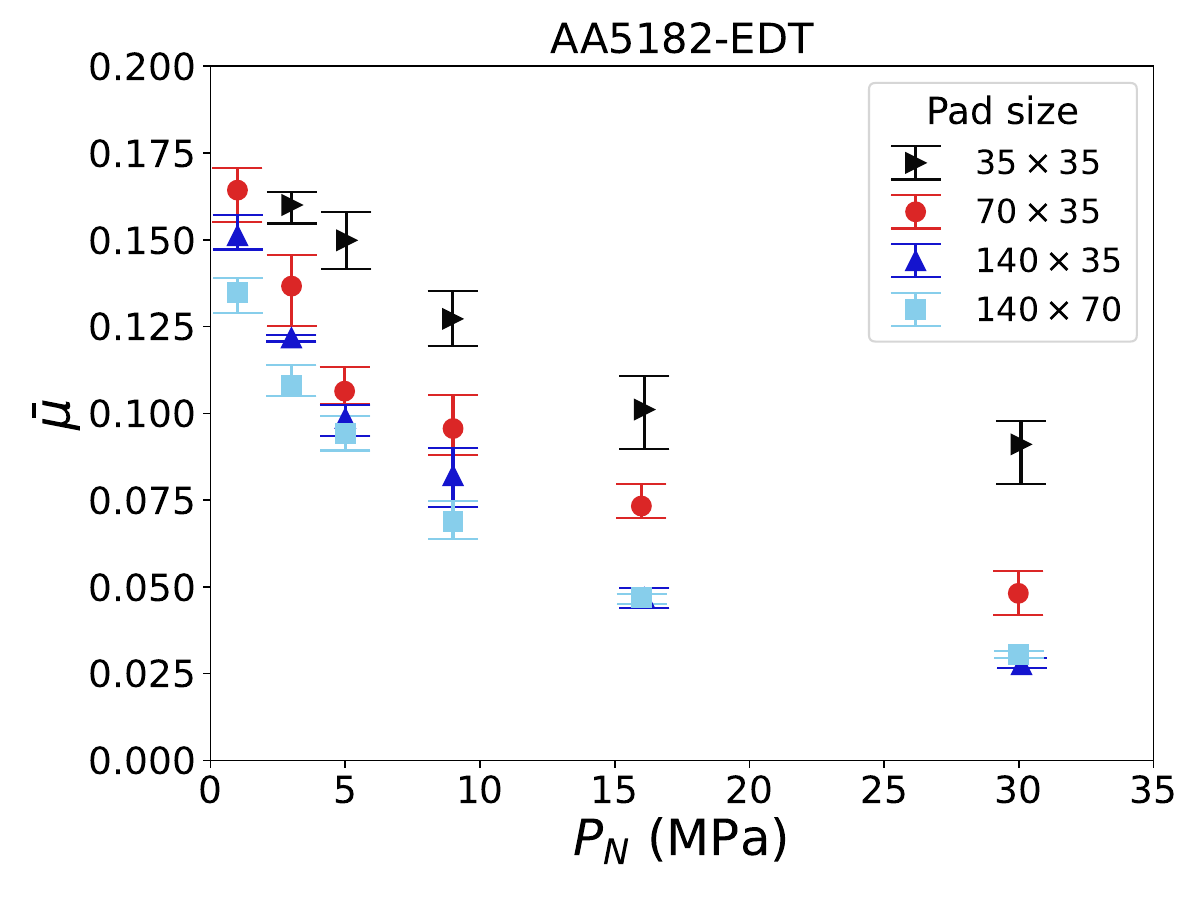}
    b\includegraphics[width=0.45\textwidth]{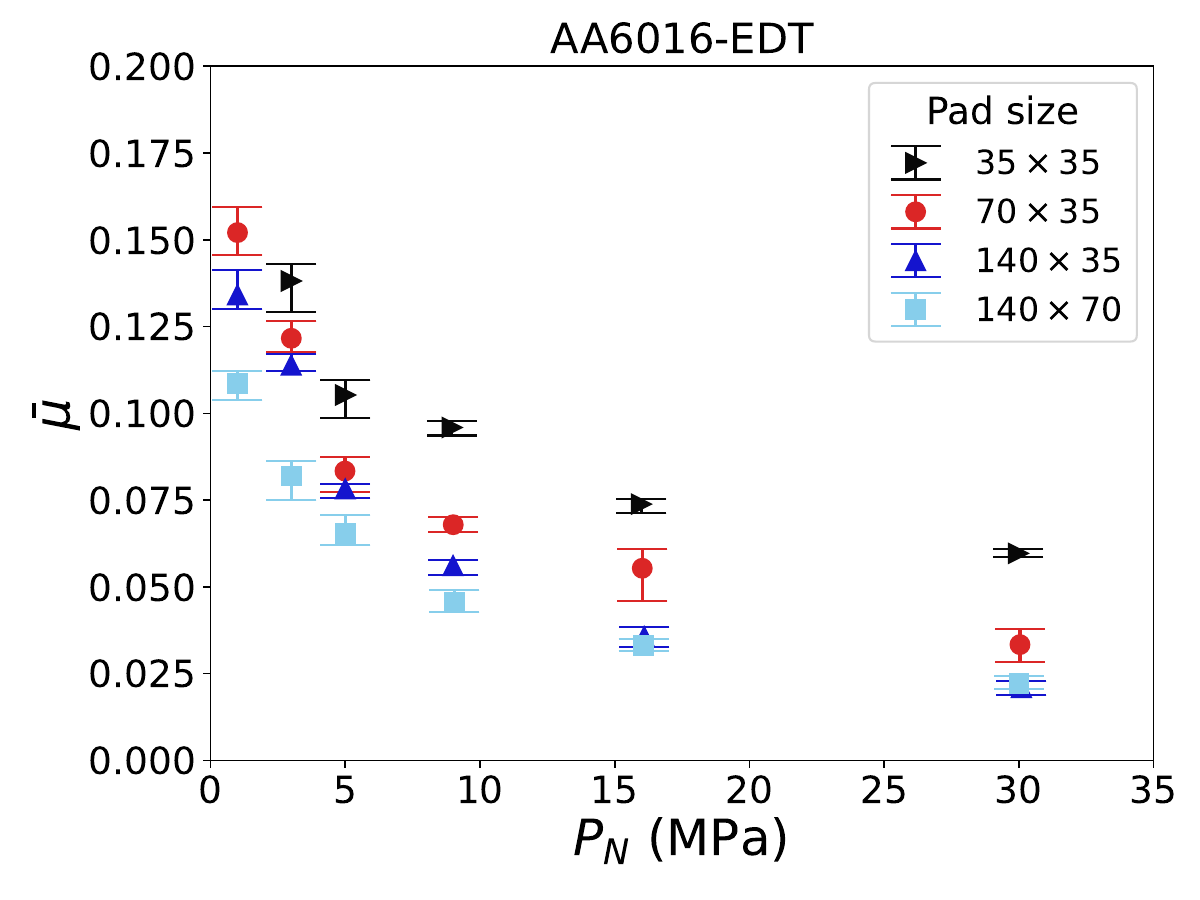}
	\caption{Evolution of the average friction coefficient, denoted $\bar{\mu}$, over the converged portion of $\mu(t)$, as a function of applied normal pressure $P_N$ for different tool-pad sizes:  
(a) AA5182 alloy sheet specimens in the O-temper condition with an EDT surface finish;  
(b) AA6016 alloy sheet specimens in the T4-temper condition with an EDT surface finish.
}
	\label{fig:exp-width}
\end{figure}

The experimental results of the friction coefficients measured on AA6016 and AA5054 EDT, shown in Figure~\ref{fig:exp-width}, indicate that the 140~$\times$~70 mm and 140~$\times$~35 mm pads exhibited nearly identical friction coefficients under the same applied pressure, despite their difference in width. In contrast, the 70~$\times$~35 mm and 35~$\times$~35 mm pads showed clear differences, suggesting a stronger dependence on pad length rather than width. This trend is consistent with observations from repeated sliding tests, indicating that tribological behavior, particularly the evolution of the friction coefficient, is strongly influenced by the sliding length of the contact interface. These findings highlight the role of tribolayer formation and its directional dependence in governing frictional performance during sliding.

\subsection{FEM formulation of the 1D macro model}
\label{sec:Macro-formulation}

This appendix details the solving scheme of the one‑dimensional bar described in Section \ref{sec:macro-num-model}.  
To derive the weak form, we multiply the governing equation \eqref{eq:macro_u_corrected} by a virtual displacement field $\delta U(x)$ and integrate over the bar domain $\Omega = [0, L_B]$, where $L_B$ is the bar length:
\begin{equation}
\int_{\Omega} \rho\, \ddot{U}(x,t)\, \delta U(x)\, dx
+ \int_{\Omega} E^*\, \frac{\partial U(x,t)}{\partial x} \, \frac{\partial \delta U(x)}{\partial x} \, dx
= \int_{\Gamma_c} T_f(x,t)\, \delta U(x)\, dx.
\end{equation}

Here, $T_f(x,t)$ denotes the tangential traction (frictional stress) exerted on the bar by the rigid counterface, and $\Gamma_c$ represents the portion of the interface where the bar is in contact with the tool.

In the finite element implementation, the displacement field $U(x,t)$ and the virtual displacement field $\delta U(x)$ are approximated using piecewise continuous shape functions. This leads to a semi-discrete system of second-order ordinary differential equations \cite{Logan1997}:
\begin{equation}
\mathbf{M}\, \ddot{\mathbf{U}}(t) + \mathbf{K}\, \mathbf{U}(t) = \mathbf{f}_{\mathrm{ext}}(t),
\label{eq:FE_discrete}
\end{equation}
where $\mathbf{U}(t)$ is the vector of nodal displacements, $\mathbf{M}$ is the consistent mass matrix, $\mathbf{K}$ is the global stiffness matrix, and $\mathbf{f}_{\mathrm{ext}}(t)$ is the vector of external forces, including contact and friction contributions.

Equation \eqref{eq:FE_discrete} is integrated in time using a predictor–corrector
scheme equivalent to the explicit central difference method (Newmark parameters
$\beta = 0$, $\gamma = 1/2$). The displacement, velocity, and acceleration at
time step $t_{n+1} = t_n + \Delta t$ are updated as:
\begin{align}
\mathbf{U}_{n+1} &= \mathbf{U}_n + \Delta t\, \mathbf{\dot{U}}_n
+ \tfrac{\Delta t^2}{2}\, \mathbf{\ddot{U}}_n, \\
\mathbf{\dot{U}}_{n+1} &= \mathbf{\dot{U}}_n + \tfrac{\Delta t}{2}\,
\left(\mathbf{\ddot{U}}_n + \mathbf{\ddot{U}}_{n+1}\right).
\end{align}

The predictor–corrector algorithm proceeds as follows:
\begin{itemize}
  \item[] \textbf{Predictor step:} estimate $\mathbf{U}_{n+1}^{\text{pred}}$
  and $\mathbf{\dot{U}}_{n+1}^{\text{pred}}$ from known values at step $n$;
  \item[] \textbf{Force computation:} evaluate internal and contact/friction
  forces based on the predicted configuration;
  \item[] \textbf{Corrector step:} solve the mass matrix system to obtain the
  accelerations $\mathbf{\ddot{U}}_{n+1}$ and update the state accordingly.
\end{itemize}
This explicit scheme enables efficient simulation of contact–friction problems with evolving local conditions, including history-dependent friction behavior.

\subsection{Derivation of the particle source term from Archard's law} \label{sec:Archard}
The rate of production of third-body particles due to wear can be derived from Archard's classical wear law, which expresses the wear volume generation rate per unit time, $\frac{dW}{dt}$, as \cite{Archard1953}:
\begin{equation}
	\frac{dW}{dt} = K  \frac{F_N}{H} |\dot{U}|,
\end{equation}
where $K$ is the dimensionless Archard wear coefficient, $F_N$ is the normal load, and $H$ is the hardness of the softer material. To obtain a local formulation based on the contact pressure $P$ and apparent contact area $A_0$, we use $F_N = P A_0$, leading to:
\begin{equation}
	\frac{dW}{dt} = K  \frac{P A_0}{H}  |\dot{U}|.
	\label{eq:dWdt}
\end{equation}
Our aim is to use the above expression to determine the rate of particle generation in the form of $\frac{d\beta}{dt}$, to be used in the state evolution Equation \ref{eq:beta_evolution}. The time derivative of $\beta$ can be expressed as:
\begin{equation}
	\frac{d\beta}{dt} = \frac{a_p \dot{N}_p}{A_0},
	\label{eq:db}
\end{equation}
where $a_p$ is the projected area of a single particle and $\dot{N}_p$ is the rate of particle generation.
Assuming that the wear volume rate $\frac{dW}{dt}$ is fully converted into particles of constant volume $v_p = a_p h_p$, the particle generation rate is:
\begin{equation}
	\dot{N}_p = \frac{dW/dt}{v_p} = \frac{1}{a_p h_p}  \frac{dW}{dt}.
\end{equation}
Substituting into Equation \ref{eq:db}, the source term in the state evolution equation (right-hand side of Equation \ref{eq:beta_evolution}) becomes:
\begin{equation}
	\frac{d\beta}{dt} = \frac{1}{h_p A_0}  \frac{dW}{dt} = K\frac{P}{H} \frac{|\dot{U}|}{h_p}.
\end{equation}

\subsection{Comparison between BEM and analytical model for real contact area}
\label{sec:Ac-comparision}
\begin{figure}[H]
	\centering
	\includegraphics[width=0.55\textwidth]{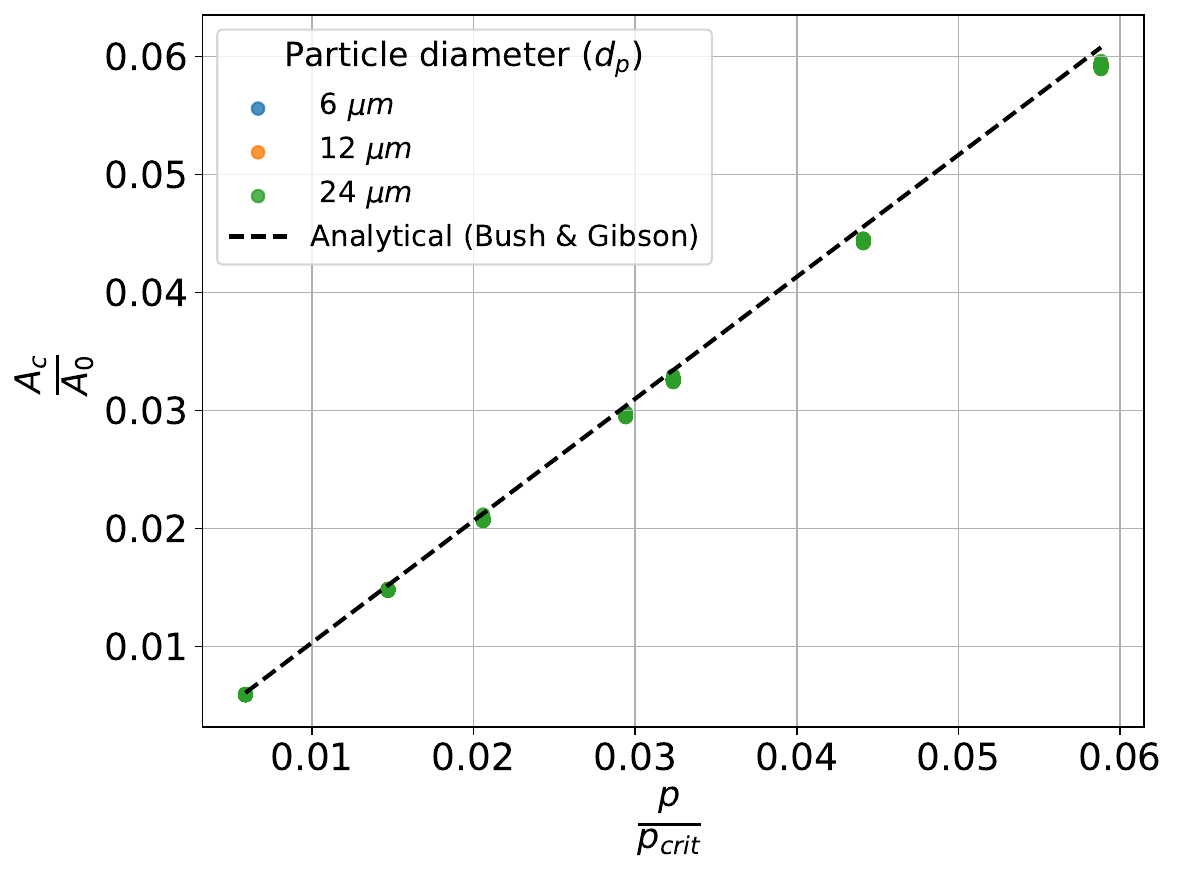}
	\caption{The figure shows a comparison between the contact area obtained from mesoscale BEM simulations, performed for three different particle diameters between rough surfaces, and the analytical prediction from the Bush and Gibson model \cite{Bush}. Due to identical values, the simulation points overlap across all sliding distances and particle diameters.}
	\label{fig:Ac_cpmparision}
\end{figure}

Figure \ref{fig:Ac_cpmparision} presents a comparison between the real contact area obtained from mesoscale BEM simulations and the analytical prediction given by the Bush and Gibson model \cite{Bush}, as expressed in Equation \ref{eq:Ac}. The simulations were conducted for three different particle diameters placed between rough surfaces. As the BEM results closely follow the analytical trend, all simulated points lie on top of the analytical curve. Due to this overlap, individual data points are not distinguishable in the figure.

%
%

\subsection{Contact nucleation phase} \label{sec:nucleation} 

When the area of each contact cluster is negligible compared to the total real contact area, the distance between clusters can be approximated by the distance between their centers.
In this scenario, the primary mechanism increasing the total contact area is the nucleation of new clusters. Knowing that the contact area is proportional to the normal pressure \cite{Bush,molinari}, we can assume that the number of clusters per unit area, $N_c$, is also proportional to the normal pressure (i.e., $N_c \propto P$).

Consider $N_c$ clusters uniformly distributed within a two-dimensional periodic square domain of side length $L$.
For sufficiently large $N_c$, this uniform distribution approximates a Poisson point process with intensity $\gamma $, where $\gamma$ is the contact cluster density, written as:
\begin{equation}
	\gamma = \frac{N_c}{L^2}.
\end{equation}

Our goal is to derive the characteristic average distance between clusters, $\bar{d_c}$. For a Poisson process in two dimensions, the probability of having no contact cluster within a radius $r$ around a reference cluster is \cite{Clark,kendall_book}:
\begin{equation}
	\mathcal{P}(r) = e^{-\gamma \pi r^2}.
\end{equation}

Differentiating this with respect to $r$ gives the probability density function (PDF) for the nearest-neighbor distance:
\begin{equation}
	f(r) = \frac{d}{dr}[1 - e^{-\gamma \pi r^2}] = 2\pi\gamma r e^{-\gamma \pi r^2}.
\end{equation}

The expected nearest-neighbor distance $d_c$ is obtained by integrating over all distances:
\begin{equation}
	\bar{d_c} = \int_0^\infty r f(r) \, dr = \int_0^\infty 2\pi\gamma r^2 e^{-\gamma \pi r^2} \, dr.
\end{equation}

Evaluating this integral yields:
\begin{equation}
	\bar{d_c} = \frac{1}{\sqrt{\pi\gamma}}.
\end{equation}

Substituting back the cluster density $ \gamma = \frac{N_c}{L^2} $, we obtain:
\begin{equation}
	\bar{d_c} = \frac{L}{\sqrt{\pi N_c}}.
	\label{eq:dc_nucleation}
\end{equation}

Since $ N_c \propto p $, $\bar{d_c}$ can be expressed as:
\begin{equation}
	\bar{d_c} \propto \frac{1}{\sqrt{p}}.
\end{equation}

This relationship describes a scenario in which new contact clusters nucleate randomly and remain small, so their individual sizes have a minimal influence on the spacing.

\subsection{Contact growth phase }\label{sec:growth} 

Assuming that the number of contact clusters, $N_c$, remains constant, the total real contact area increases with pressure due to the linear growth in the size of each cluster.
Consider each contact cluster having an average area $\bar{a_c}$, growing linearly with normal pressure $ p $. Since each cluster's radius expands with pressure for a contact cluster with an arbitrary shape, the average border-to-border distance decreases as follows:
\begin{equation}
	\bar{d_c}= \bar{d} - 2\gamma' \sqrt{\bar{a_c}}.
\end{equation}
where  $\bar{d}$ is the average center-to-center distance between clusters, and $\gamma'$ is a shape-dependent geometric factor.
Because the real contact area grows with cluster size and scales approximately linearly with the normal pressure $p$, we set $\bar{a_c} \propto p$.
Consequently,
\begin{equation}
	\bar{d_c} \;\propto\; \bar{d} - C\sqrt{p},
	\label{eq:dc_growth}
\end{equation}
where $ C $ is a geometrical constant that depends on $\gamma'$.

Here, we have only considered two separate cases: the nucleation and growth of contact clusters. In reality, however, both processes occur simultaneously.
Nevertheless, it is still worthwhile to examine each mechanism independently to gain a clearer understanding.
Moreover, if contact clusters grow large enough, they may eventually merge. This merging is not considered in the current analysis but could occur at higher normal pressures beyond the scope of this study.

\newpage

\section*{Author Contributions}

\textbf{P.A.} developed the numerical models (excluding the 1D macroscale FEM code), contributed to the conceptualization of the study, and co-wrote the original draft.  
\\
\textbf{G.A.} implemented the 1D macroscale FEM code, contributed to the conceptualization, co-wrote the original draft, and supervised the project together with J.-F.M.  
\\
\textbf{J.-F.M.} contributed to the conceptualization and supervised the project.  
\\
\textbf{L.R.} performed the experimental studies, and co-wrote the original draft.  
\\
\textbf{C.L.} was the project manager.   
\\
All authors contributed to the review and editing of the manuscript.

\newpage

\addcontentsline{toc}{section}{References}
\bibliographystyle{ieeetr}
\bibliography{references}

\end{document}